\documentclass{JHEP3}

\usepackage{amsmath}
\usepackage{graphicx}
\usepackage{url}
\usepackage{epsfig}
\usepackage{cite}

\newcommand{\sq}{{\tilde{q}}}
\newcommand{\sqb}{{\bar{\tilde{q}}}}
\newcommand{\gl}{{\tilde{g}}}
\renewcommand{\d}{\mathrm{d}}
\newcommand{\nn}{\nonumber}

\title{NNLL resummation for squark-antisquark pair production at the LHC}

\author{Wim Beenakker\\
  Theoretical High Energy Physics, IMAPP, Faculty of Science, Mailbox 79, Radboud University Nijmegen, P.O. Box 9010, NL-6500 GL Nijmegen, The Netherlands}
  \author{Silja Brensing\\
  DESY, Theory Group, Notkestrasse 85, D-22603 Hamburg, Germany}
\author{Michael Kr\"amer, Anna Kulesza\\
  Institute for Theoretical Particle Physics and Cosmology, RWTH Aachen University\\
  D-52056 Aachen, Germany}
\author{Eric Laenen\\
  ITFA, University of Amsterdam, Valckenierstraat 65, 1018 XE Amsterdam, \\
  ITF, Utrecht University, Leuvenlaan 4, 3584 CE Utrecht\\
  Nikhef Theory Group, Science Park 105, 1098 XG Amsterdam, The
  Netherlands}
\author{ Irene Niessen\\
 Theoretical High Energy Physics, IMAPP, Faculty of Science Mailbox, 79, Radboud University Nijmegen, P.O. Box 9010, NL-6500 GL Nijmegen, The Netherlands}

\abstract{We consider the resummation of soft gluon emission for squark-antisquark pair-production at the LHC at next-to-next-to-leading-logarithmic (NNLL) accuracy in the framework of the minimal supersymmetric standard model. We present the analytical ingredients needed for the calculation and provide numerical predictions for the LHC at centre-of-mass energies of 7 and 14 TeV. We find a significant reduction in the scale uncertainty and a considerable increase in the prediction of the total cross section. Compared to the next-to-leading order prediction, the corrections increase the cross section by up to 30\% for 1.5~TeV squarks at a centre-of-mass energy of 7~TeV.}


\preprint{DESY 11-175\\ ITP-UU-11/36\\ ITFA 11-14\\ NIKHEF/2011-027\\ TTK-11-47}

\begin{document}

\section{Introduction}
\label{s:intro}

The search for supersymmetry (SUSY) \cite{Golfand:1971iw,Wess:1974tw} is a central part of the experimental program at the Large Hadron Collider (LHC). Models of weak-scale SUSY provide a promising solution to the hierarchy problem of the Standard Model (SM) and predict new supersymmetric particles (sparticles) with masses in the TeV range. If they exist, the coloured sparticles, squarks ($\tilde q$) and gluinos ($\tilde g$), would be produced copiously in hadronic collisions and thus offer the strongest sensitivity for supersymmetry searches at the LHC. In the context of the minimal supersymmetric extension of the Standard Model (MSSM) \cite{Nilles:1983ge,Haber:1984rc}, these particles are produced in pairs due to $R$-parity conservation. 

Searches for squarks and gluinos at the proton-proton collider LHC, which has been operating at $\sqrt{S}=7$~TeV in 2010 and 2011, have placed lower limits on squark and gluino masses around 1~TeV \cite{Aad:2011ib,Chatrchyan:2011zy}. Once the LHC reaches its design energy of $\sqrt{S}=14$~TeV, SUSY particles with masses up to 3~TeV can be probed \cite{Aad:2009wy,Bayatian:2006zz}.

Accurate theoretical predictions for inclusive squark and gluino cross sections are needed both to set exclusion limits and, in case SUSY is discovered, to determine SUSY particle masses and properties \cite{Baer:2007ya,Dreiner:2010gv}. The inclusion of higher-order SUSY-QCD corrections significantly reduces the renormalization- and factorization-scale dependence of the predictions.  In general, the corrections also increase the size of the cross section with respect to the leading-order prediction \cite{Kane:1982hw,Dawson:1983fw} if the renormalization and factorization scales are chosen close to the average mass of the produced SUSY particles. Consequently, the SUSY-QCD corrections have a substantial impact on the determination of mass exclusion limits and would lead to a significant reduction of the uncertainties on SUSY mass or parameter values in the case of discovery.  The squark-antisquark production processes have been known for quite some time at next-to-leading order (NLO) in SUSY-QCD \cite{Beenakker:1994an,Beenakker:1996ch}. Electroweak corrections to the ${\cal O} (\alpha_{\rm s}^2)$ tree-level processes \cite{Hollik:2008yi} and the electroweak Born production channels of ${\cal O} (\alpha\alpha_{\rm s})$ and ${\cal O} (\alpha^2)$ \cite{Bornhauser:2007bf} are in general significant for the pair production of SU(2)-doublet squarks $\tilde{q}_L$ and at large invariant masses, but they are moderate for inclusive cross sections.

A significant part of the NLO QCD corrections can be attributed to the threshold region, where the partonic centre-of-mass energy is close to the kinematic production threshold. In this region the NLO corrections are dominated by soft-gluon emission off the coloured particles in the initial and final state and by the Coulomb corrections due to the exchange of gluons between the slowly moving massive sparticles in the final state. The soft-gluon corrections can be taken into account to all orders in perturbation theory by means of threshold resummation techniques \cite{Sterman:1986aj,Catani:1989ne}. The Coulomb corrections can be summed to all orders by either using a Sommerfeld factor \cite{BBPC:BBPC19400460520,Sakharov:1948yq,Catani:1996dj,Fadin:1990wx} or by employing the framework of non-relativistic QCD, where bound-state effects can be included as well \cite{Fadin:1990wx,Kiyo:2008bv,Hagiwara:2008df,Hagiwara:2009hq,Kauth:2011bz,Kauth:2011vg}. In addition, a formalism has been developed in the framework of effective field theories that allows for the combined resummation of soft and Coulomb gluons in the production of coloured sparticles \cite{Beneke:2009rj,Beneke:2010da}.

Threshold resummation has been performed for all MSSM squark and gluino production processes at next-to-leading-logarithmic (NLL) accuracy \cite{Kulesza:2008jb,Kulesza:2009kq,
Beenakker:2009ha,Beenakker:2010nq,Beenakker:2011fu}. For squark-antisquark production, in addition to soft-gluon resummation, the Coulomb corrections have been resummed both by using a Sommerfeld factor \cite{Kulesza:2009kq} and by employing the framework of effective field theories \cite{Beneke:2010da}. Furthermore, the
dominant next-to-next-to-leading order (NNLO) corrections, including those coming from the resummed cross section at
next-to-next-to-leading-logarithmic (NNLL) level, have been calculated for squark-antisquark pair-production \cite{Langenfeld:2009eg}. 

In this paper we consider threshold resummation at  NNLL accuracy for squark-anti\-squark pair production at the LHC. Compared to the NLL calculation the new ingredients are the one-loop matching coefficients, which contain the NLO cross section near threshold, and the two-loop soft anomalous dimensions. Studies for the pair production of top quarks suggest that the effect of the matching coefficients can be significant \cite{Bonciani:1998vc} and that NNLL resummation can reduce the scale dependence considerably \cite{Kidonakis:2001nj,Langenfeld:2009eg,Kidonakis:2010dk,Ahrens:2010zv,Beneke:2011mq}. We will discuss the impact of the corrections and provide an estimate of the theoretical uncertainty due to scale variation. In addition we will study the impact of the Coulomb gluons on the cross section. We exclude top squarks from the final state in view of potentially large mixing effects and mass splitting in the stop sector~\cite{Ellis:1983ed}. The other squarks are considered as being mass degenerate and all flavours and chiralities are summed over.

The structure of the paper is as follows. In section~\ref{s:resummation} we discuss the NNLL resummation for squark-antisquark pair-production. In section~\ref{s:Ccoeff} we present the calculation of the hard matching coefficients required for the NNLL resummation. The numerical results are presented in section~\ref{s:numres}. We show predictions for the LHC with centre-of-mass energies of $\sqrt{S}=7$~TeV and $\sqrt{S}=14$~TeV. We will conclude in section~\ref{s:conclusion}.

\section{Threshold resummation at NNLL}
\label{s:resummation}

In this section we briefly review the formalism of threshold resummation for the production of a squark-antisquark pair. The inclusive hadroproduction cross section $\sigma_{h_1h_2\rightarrow \sq\sqb}$ can be written in terms of its partonic version
$\sigma_{ij\rightarrow \sq\sqb}$ as
\begin{multline}
  \label{eq:7}
  \sigma_{h_1 h_2 \to \sq\sqb}\bigl(\rho, \{m^2\}\bigr) 
  \;=\; \sum_{i,j} \int d x_1 d x_2\,d\hat{\rho}\;
        \delta\left(\hat{\rho} - \frac{\rho}{x_1 x_2}\right)\\
        \times\,f_{i/h_{1}}(x_1,\mu^2 )\,f_{j/h_{2}}(x_2,\mu^2 )\,
        \sigma_{ij \to \sq\sqb}\bigl(\hat{\rho},\{ m^2\},\mu^2\bigr)\,,
\end{multline}
where $\{m^2\}$ denotes all masses entering the calculations, $i,j$ are the initial parton flavours, $f_{i/h_1}$ and $f_{j/h_2}$ the parton distribution functions and $\mu$ is the common factorization and renormalization scale. The hadronic threshold for inclusive production of two final-state squarks with mass $m_\sq$ corresponds to a hadronic centre-of-mass energy squared that is equal to $S=4m_\sq^2$. Thus we define the threshold variable $\rho$, measuring the distance from threshold in terms of energy fraction, as
\[\rho \;=\; \frac{4m_\sq^2}{S}\,.\]
The partonic equivalent of this threshold variable is defined as $\hat{\rho}=\rho/(x_1x_2)$, where $x_{1,2}$ are the momentum fractions of the partons. 

In the threshold region, the dominant contributions to the higher-order QCD corrections due to soft-gluon emission have the general form
\begin{equation}
\alpha_{\rm s}^n \log^m\!\beta^2\ \ , \ \ m\leq 2n 
\qquad {\rm \ with\ } \qquad 
\beta^2 \,\equiv\, 1-\hat{\rho} \,=\, 1 \,-\, \frac{4m_\sq^2}{s}\,,
\label{eq:beta}
\end{equation}
where $s=x_1x_2S$ is the partonic centre-of-mass energy squared. The resummation of the soft-gluon contributions is performed after taking a Mellin transform (indicated by a tilde) of the cross section,
\begin{align}
  \label{eq:10}
  \tilde\sigma_{h_1 h_2 \to \sq\sqb}\bigl(N, \{m^2\}\bigr) 
  &\equiv \int_0^1 d\rho\;\rho^{N-1}\;
           \sigma_{h_1 h_2\to \sq\sqb}\bigl(\rho,\{ m^2\}\bigr) \nonumber\\
  &=      \;\sum_{i,j} \,\tilde f_{i/{h_1}} (N+1,\mu^2)\,
           \tilde f_{j/{h_2}} (N+1, \mu^2) \,
           \tilde{\sigma}_{ij \to \sq\sqb}\bigl(N,\{m^2\},\mu^2\bigr)\,.
\end{align}
The logarithmically enhanced terms are then of the form $\alpha_{\rm
  s}^n \log^m N$, $m\leq 2n$, with the threshold limit
$\beta\rightarrow 0$ corresponding to $N\rightarrow \infty$. 
The all-order summation of such logarithmic terms is a consequence of the
near-threshold factorization of the cross sections into functions that
each capture the contributions of classes of radiation effects: hard,
collinear (including soft-collinear), and wide-angle soft radiation
\cite{Sterman:1986aj,Catani:1989ne,Bonciani:1998vc,Contopanagos:1996nh,
  Kidonakis:1998bk,Kidonakis:1998nf}.
Near threshold the resummed partonic cross section has the form:

\begin{align}
  \label{eq:12}
  \tilde{\sigma}^{\rm (res)} _{ij\rightarrow \sq\sqb}\bigl(N,\{m^2\},&\mu^2\bigr) 
  =\sum_{I}\,
      \tilde\sigma^{(0)}_{ij\rightarrow \sq\sqb,I}\bigl(N,\{m^2\},\mu^2\bigr)\, 
C_{ij\rightarrow \sq\sqb,I}(N,\{m^2\},\mu^2)\nonumber\\
  & \times\,\Delta_i (N+1,Q^2,\mu^2)\,\Delta_j (N+1,Q^2,\mu^2)\,
     \Delta^{\rm (s)}_{ij\rightarrow \sq\sqb,I}\bigl(Q/(N\mu),\mu^2\bigr)\,,
\end{align}
where we have introduced the hard scale $Q^2 = 4m_\sq^2$. Before commenting on the different functions in this equation, we recall that soft radiation
is coherently sensitive to the colour structure of the hard process
from which it is emitted \cite{Bonciani:1998vc,Contopanagos:1996nh,Kidonakis:1998bk,Kidonakis:1998nf,Botts:1989kf,Kidonakis:1997gm}. At threshold, the resulting colour matrices become diagonal to all orders by performing the calculation in an $s$-channel colour basis \cite{Beneke:2009rj,Kulesza:2008jb,Kulesza:2009kq}. The different contributions then correspond to different irreducible representations~$I$. For the $q\bar q\to\sq\sqb$ process, the $s$-channel basis consists of a singlet $\bf1$ and an octet $\bf8$ representation, while for the $gg\to\sq\sqb$ process it contains a singlet $\bf1$, an antisymmetric octet $\bf8_A$ and a symmetric octet $\bf8_S$ representation as presented in e.g. Ref.~\cite{Beneke:2009rj,Kulesza:2008jb,Kulesza:2009kq}.
 
In Eq.~\eqref{eq:12}, $\tilde{\sigma}^{(0)}_{ij \rightarrow \sq\sqb,I}$ are the colour-decomposed leading-order (LO) cross sections in Mellin-moment space. The functions $\Delta_{i}$ and $\Delta_{j}$ sum the effects of the (soft-)collinear radiation from the incoming partons, while the function $\Delta^{\rm (s)}_{ij\rightarrow \sq\sqb,I}$ describes the wide-angle soft radiation. Schematically the exponentiation of soft-gluon radiation takes the form \cite{Sterman:1986aj,Catani:1989ne}
\begin{equation}
  \label{eq:3}
\Delta_i\Delta_j\Delta^{\rm(s)}_{ij\rightarrow \sq\sqb,I}
  \;=\; \exp\Big[L g_1(\alpha_{\rm s}L) + g_2(\alpha_{\rm s}L) + \alpha_{\rm s}g_3(\alpha_{\rm s}L) + \ldots \Big]  \,.
\end{equation}
This exponent captures all dependence on the large logarithm $L=\log N$. Keeping only the $g_1$ term in Eq.~\eqref{eq:3} constitutes the leading logarithmic (LL) approximation. Including also the $g_2$ term is called the NLL approximation. For the NNLL approximation also the $g_3$ term needs to be taken into account. Explicit expressions for the $g_3$ term and its ingredients are given in Refs.~\cite{Moch:2005ba,Czakon:2009zw,Moch:2008qy}\footnote{One has to correct for an extra minus sign in front of all $D_{Q\bar Q}$ terms in Eq.~(A9) of \cite{Moch:2008qy}.} and are listed in appendix~\ref{app:g3}.

We also need the matching coefficients $C$, which contain the Mellin moments of the higher-order contributions without the $\log(N)$ terms. To NNLL accuracy, this non-logarithmic part of the higher-order cross section near threshold factorizes into a part that contains the leading Coulomb correction ${\cal C}^{\rm Coul,(1)}$ and a part that contains the NLO hard matching coefficients ${\cal C}^{\rm (1)}$ \cite{Beneke:2010da}:
\begin{align}
C^{\rm NNLL}=\left(1+\frac{\alpha_{\rm s}}{\pi}\;{\cal C}^{\rm Coul,(1)}(N,\{m^2\},\mu^2)\right)\;\left(1+\frac{\alpha_{\rm s}}{\pi}\;{\cal C}^{\rm (1)}(\{m^2\},\mu^2)\right)\label{eq:matchingcoeff}
\end{align}
The calculation of the NLO hard matching coefficients ${\cal C}^{\rm (1)}$ and the Coulomb contribution ${\cal C}^{\rm Coul,(1)}$ for the squark-antisquark production processes will be discussed in detail in section~\ref{s:Ccoeff}.

Having constructed the NNLL cross-section in Mellin-moment space, the inverse Mellin transform has to be performed in order to recover the hadronic cross section $\sigma_{h_1 h_2 \to \sq\sqb}$. In order to retain the information contained in the complete NLO cross sections~\cite{Beenakker:1996ch}, the NLO and NNLL results are combined through a matching procedure that avoids double counting of the NLO terms:
\begin{align}
  \label{eq:matching}
  \sigma^{\rm (NLO+NNLL~matched)}_{h_1 h_2 \to \sq\sqb}&\bigl(\rho, \{m^2\},\mu^2\bigr) 
  =\; \sigma^{\rm (NLO)}_{h_1 h_2 \to \sq\sqb}\bigl(\rho, \{m^2\},\mu^2\bigr)
          \\[1mm]
& +\, \sum_{i,j}\,\int_\mathrm{CT}\,\frac{dN}{2\pi i}\,\rho^{-N}\,
       \tilde f_{i/h_1}(N+1,\mu^2)\,\tilde f_{j/h_{2}}(N+1,\mu^2) \nonumber\\[2mm]
&\times\,
       \left[\tilde\sigma^{\rm(res,NNLL)}_{ij\to \sq\sqb}\bigl(N,\{m^2\},\mu^2\bigr)
             \,-\, \tilde\sigma^{\rm(res,NNLL)}_{ij\to \sq\sqb}\bigl(N,\{m^2\},\mu^2\bigr)
       {\left.\right|}_{\scriptscriptstyle{\rm (NLO)}}\, \right]. \nonumber
\end{align}
We adopt the ``minimal prescription'' of Ref.~\cite{Catani:1996yz} for
the contour CT of the inverse Mellin transform in Eq.~(\ref{eq:matching}).
In order to use standard parametrizations of parton distribution
functions in $x$-space we employ the method introduced in
Ref.~\cite{Kulesza:2003wn}.

\section{Calculation of the matching coefficients}
\label{s:Ccoeff}

In this section we will discuss the calculation of the matching coefficients $C$ at one loop. As discussed in equation~\eqref{eq:matchingcoeff}, the NNLL matching coefficient $C^{\rm NNLL}$ factorizes into the Coulomb contribution and the hard matching coefficient. For NNLL resummation, both terms are needed up to NLO accuracy.

The terms in the NLO cross section which give rise to the Coulomb
corrections ${\cal C}^{\rm Coul,(1)}$ in $N$-space do not have the
usual phase-space suppression~\mbox{$\propto\beta$}, in view of the Coulombic
{$1/\beta$} enhancement factor. After performing an expansion of the
NLO cross section in $\beta$, the hard matching coefficients ${\cal
  C}^{\rm(1)}$  are determined by the terms in the NLO cross section
that are proportional to $\beta$,  $\beta \log(\beta)$ and  $\beta \log^2(\beta)$. Terms that contain higher powers of $\beta$ are suppressed by powers of $1/N$ in Mellin-moment space and do not contribute to the matching coefficient $C$. In contrast to the case of top-pair production in Ref.~\cite{Czakon:2008cx}, there is no full analytic result for the real corrections to squark-antisquark production, so we cannot take the explicit threshold limit. For the virtual corrections, which also contain the Coulomb contribution, we will use the full analytic expressions, but for the real corrections we need a different approach.

To obtain the virtual corrections for squark-antisquark production we start from the full analytic calculation as presented in Ref.~\cite{Beenakker:1996ch}. As described in detail in Ref.~\cite{Beenakker:1996ch}, the QCD coupling $\alpha_{\rm s}$ and the parton distribution functions at NLO are defined in the $\overline{\rm MS}$ scheme with five active flavours, with a correction for the SUSY breaking in the $\overline{\rm MS}$ scheme. The masses of squarks and gluinos are renormalized in the on-shell scheme, and the top quark and the SUSY particles are decoupled from the running of $\alpha_{\rm s}$.

To obtain the virtual part of the hard matching coefficients, we first need to colour-decompose the result and then expand it in $\beta$. For the first step we only need the colour decomposition of the LO matrix element. Due to the orthogonality of the $s$-channel colour basis, the full matrix element squared is then automatically colour-decomposed:
\[|{\cal M}|^2_{{\rm NLO},I}=2\mathrm{Re}({\cal M}_{\rm NLO}{\cal M}^*_{{\rm LO},I}).\]

We are now left with an expression in terms of masses, Mandelstam variables and scalar integrals. Since we need the cross section to ${\cal O}(\beta)$, we have to expand $|{\cal M}|^2$ to zeroth order in $\beta$. For the squark-antisquark production processes the factors that multiply the integrals do not contain negative powers of $\beta$, so we do not have to expand the scalar integrals beyond zeroth order in $\beta$.

The number of integrals that need to be expanded can be reduced. By using the fact that the two outgoing momenta are equal at threshold, we can reduce some of the three- and four-point integrals to two- and three-point integrals respectively. This procedure can be used only for integrals that contain both outgoing momenta. The result of the remaining integrals is explicitly expanded to zeroth order in $\beta$.

Special attention has to be paid to the Coulomb integrals. First, in
order to calculate the Coulomb corrections ${\cal C}^{\rm Coul,(1)}$
in $N$-space, we need to know the Coulomb part of the NLO correction in
$\beta$-space, corresponding to the leading terms in $\beta$ of the
Coulomb integrals. These leading terms are given by~\cite{BBPC:BBPC19400460520,Sakharov:1948yq,Catani:1996dj,Fadin:1990wx}:

\begin{align}
\sigma_{ij\rightarrow \sq\sqb,I}^{\rm Coul,(1)}=-\frac{\alpha_{\rm s}}{\pi}\;\frac{\pi^2}{2\beta}\kappa_{ij\rightarrow \sq\sqb,I}\sigma_{ij\rightarrow \sq\sqb,I}^{\rm(0)}\label{eq:coulomb}
\end{align}
with $\kappa$ colour coefficients that depend on the process and the dimension of the representation. For the $q\bar q$-initiated process they are given by \cite{Kulesza:2009kq}:
\[\kappa_{q\bar q\to\sq\sqb,\bf1}=-\frac{4}{3} \quad\mathrm{and}\quad \kappa_{q\bar q\to\sq\sqb,\bf8}=\frac{1}{6}\] 
while for the $gg$-initiated process they are:
\[\kappa_{gg\to\sq\sqb,\bf1}=-\frac{4}{3} ,\quad \kappa_{gg\to\sq\sqb,\bf8_A}=\frac{1}{6}\quad\mathrm{and}\quad \kappa_{gg\to\sq\sqb,\bf8_S}=\frac{1}{6}\;.\] 
The Mellin transform $\tilde\sigma^{\rm Coul,(1)}$ of Eq.~\eqref{eq:coulomb} is presented in appendix~\ref{app:mellin}. The function ${\cal C}^{\rm Coul,(1)}$ can be obtained by dividing $\tilde\sigma^{\rm Coul,(1)}$ by the Mellin transform of the LO cross section, which can be found in Ref.~\cite{Kulesza:2009kq}.

Secondly, the next term in the $\beta$-expansion of the Coulomb integrals contributes to the hard matching coefficients. Due to their $1/\beta$ behaviour Coulomb integrals cannot be reduced to lower-point integrals, so they need to be expanded explicitly.

To obtain the integrated real corrections at threshold, the key observation is that they are formally phase-space suppressed near threshold unless the integrand compensates this suppression. Therefore we can construct the real corrections at threshold from the singular behaviour of the matrix element squared, which can be obtained using dipole subtraction \cite{Catani:1996vz,Catani:2002hc}. We will briefly review the procedure of dipole subtraction and specify how only the singular contributions survive in the threshold limit.

Dipole subtraction makes use of the fact that the cross section can be split into three parts: a part with three-particle kinematics $\sigma^{\{3\}}$, one with two-particle kinematics $\sigma^{\{2\}}$, and a collinear counterterm $\sigma^C$ that is needed for removing the initial-state collinear singularities. These parts are well-defined in $n=4-2\epsilon$ dimensions, but their constituents diverge for $\epsilon\to0$. With the aid of an auxiliary cross section $\sigma^A$, which captures all singular behaviour, all parts are made finite and integrable in four space-time dimensions. This auxiliary cross section is subtracted from the real corrections $\sigma^{\rm R}$ at the integrand level to obtain $\sigma^{\{3\}}$ and added to the virtual corrections $\sigma^{\rm V}$, which defines $\sigma^{\{2\}}$:
\[\sigma^{\rm NLO}=\int_{3}\big[\d\sigma^{\rm R}-\d\sigma^{\rm A}\big]_{\epsilon=0}+\int_{2}\big[\d\sigma^{\rm V}+\int_1\d\sigma^{\rm A}\big]_{\epsilon=0}+\sigma^{\rm C}\equiv\sigma^{\{3\}}+\sigma^{\{2\}}+\sigma^{\rm C}\]
We will first argue that we can neglect $\sigma^{\{3\}}$. Compared to the case of two-parton kinematics, the phase space of $\sigma^{\{3\}}$ is limited by the energy of the third, radiated massless particle. Near the two-particle threshold, the maximum energy of the radiated particle, and thus the available phase space, equals $E_{\rm max}=\sqrt{s}-2m_\sq\propto\beta^2$. Since after subtracting $\sigma^A$ no divergences are left in the integrand of $\sigma^{\{3\}}$, the leading contribution of $\sigma^{\{3\}}$ is at most proportional to $\beta^2$ and can thus be neglected. This leaves us with:
\[\sigma^{\rm NLO,thr}=\sigma^{\{2\},\rm thr}+\sigma^{\rm C,\rm thr}=\sigma^{\rm V,thr}+\sigma^{\rm C,thr}+\sigma^{\rm A,thr},\]
so at threshold the real radiation is indeed completely specified by the singular behaviour contained in $\sigma^A$. In Ref.~\cite{Catani:2002hc} the general form of $\sigma^A$ is determined by summing over dipoles that correspond to pairs of ordered partons. These dipoles describe the soft and collinear radiation and reproduce the matrix element squared in the soft and collinear limits. To obtain the cross section, the dipole functions need to be integrated over phase space and in particular over the momentum fraction $x$ that is left after radiation. In the threshold limit the available phase space sets the lower bound of the $x$-integral to $1-\beta^2$, while the upper bound equals 1. Therefore we cannot get a result of ${\cal O}(\beta)$ unless the integrand diverges at $x=1$, which is the case only for soft-gluon radiation. As a result we only need to take into account the dipoles that describe gluon radiation.

Special attention has to be paid to the massive final-state dipole function. In Ref.~\cite{Catani:2002hc} this dipole function has been rewritten in order to simplify the integration. Unfortunately this results in a deformation of the phase space integration which changes exactly the finite terms that we are looking for. Therefore the expression given in Eq.~(5.16) of Ref.~\cite{Catani:2002hc} cannot be used for our calculation and we have to use the original dipole function instead. A more detailed argument can be found in appendix~\ref{app:vfactors}. The divergent part of the dipole function is completely determined by the soft limit. Using the eikonal approximation we obtain the final-state dipole function that correctly reproduces the soft limit at threshold. Its behaviour is given by:
\begin{equation}
\left.\langle{\bf V}_{gj,l}\rangle\right|_{\rm div}\propto\sum_{j}\frac{1}{p_g\cdot p_j}\sum_{l\ne j}\left(\frac{p_j\cdot p_l}{p_g\cdot p_j+p_g\cdot p_l}-\frac{1}{2}\frac{m_j^2}{p_g\cdot p_j}\right)T_j\cdot T_l\;,\label{eq:dipole}
\end{equation}
where $p_g$ is the gluon momentum and the sums run over final-state particles with momenta $p_{j,l}$, masses $m_{j,l}$ and colour charge operators $T_{j,l}$ that are defined in Ref.~\cite{Catani:2002hc}. This expression vanishes at threshold, so the final-state dipoles do not contribute in the threshold limit. For the other dipole functions and the collinear counterterm we can use the equations in \cite{Catani:2002hc} and take the threshold limit.

After combining the real and the virtual corrections the hard matching coefficients can be obtained by taking the Mellin transform and omitting the Coulomb corrections and the $\log(N)$ terms. The complete expressions for the hard matching coefficients of the squark-antisquark production processes can be found in appendix~\ref{app:Ccoeff}. Their behaviour for varying gluino mass is shown in Fig.~\ref{fig:Ccoeffplots}.
\FIGURE{
\hspace{-0.55cm}
\begin{tabular}{lll}
(a)\hspace{-0.3cm}\epsfig{file=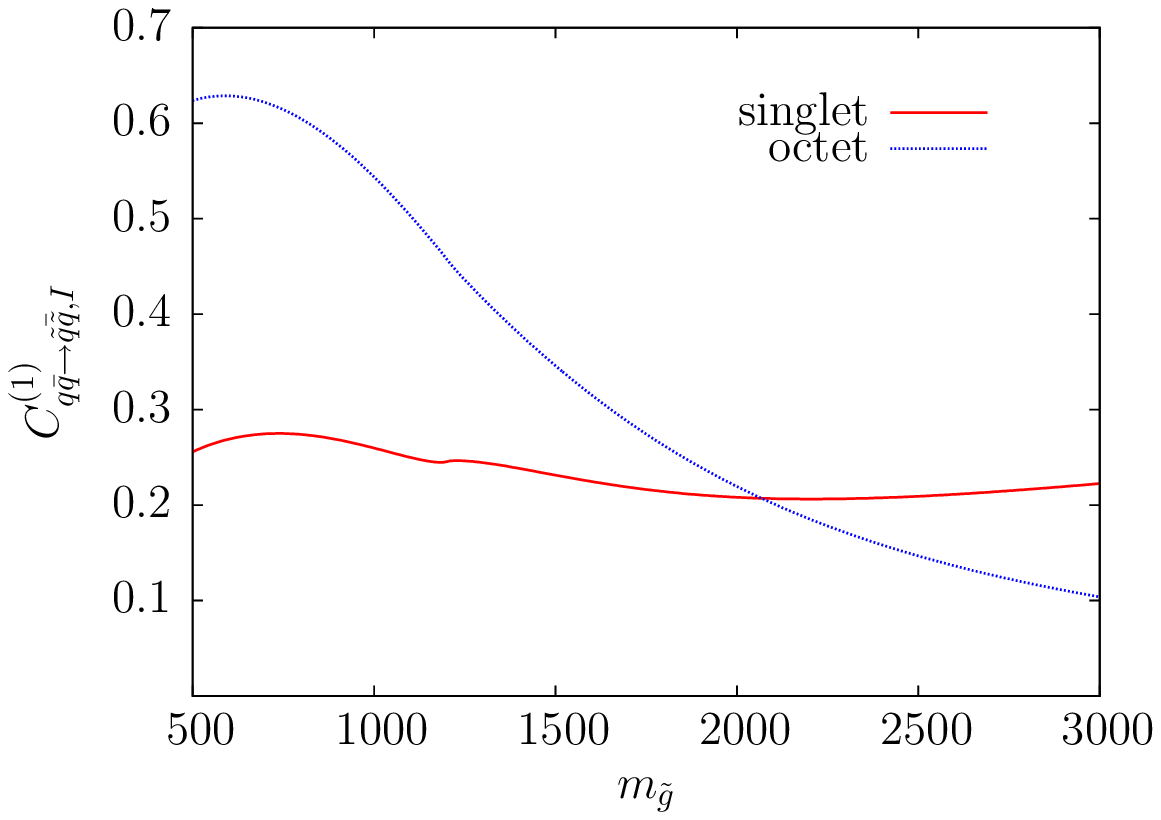, width=0.47\columnwidth}&
(b)\hspace{-0.3cm}\epsfig{file=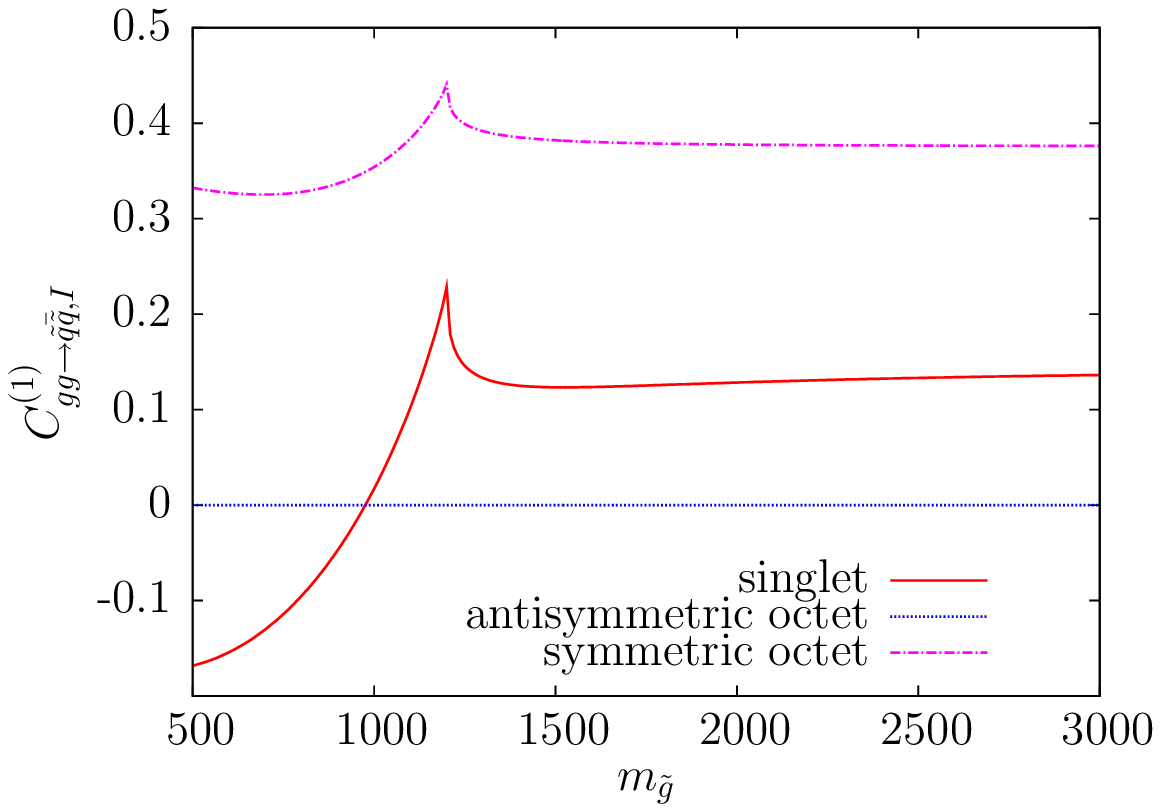, width=0.47\columnwidth}
\end{tabular}
\caption{Gluino-mass dependence of the colour-decomposed NLO hard matching coefficients for the $q\bar q$ initiated channel (a) and the $gg$ initiated channel (b). The squark mass has been set to $m_\sq=1.2$~TeV, while the common renormalization and factorization scale has been set equal to the squark mass. The top quark mass is taken to be $m_t=172.9$~GeV.\label{fig:Ccoeffplots}}
}
For the $gg\to\tilde q\bar{\tilde q}$ process the antisymmetric octet $\bf8_A$ contribution to the cross section vanishes because it yields a $p$-wave contribution, which vanishes at threshold. 

These matching coefficients have been checked numerically using {\tt PROSPINO}~\cite{Beenakker:1996ch} and agree within the numerical accuracy of {\tt PROSPINO}. They also agree with the $a_1$ coefficients presented in Ref.~\cite{Langenfeld:2009eg} to the percent level.

\section{Numerical results}
\label{s:numres}

In this section we present numerical results for the NNLL-resummed cross sections with and without the Coulomb contributions. We show the results for squark-antisquark pair-production at the LHC for centre-of-mass energies of 7~TeV and 14~TeV. In order to evaluate hadronic cross sections we use the 2008 NLO MSTW parton distribution functions~\cite{Martin:2009iq} with the corresponding $\alpha_{\rm s}(M_{Z}^2) = 0.120$. We have used a top quark mass of $m_t=172.9$~GeV~\cite{PDG}. The numerical results have been obtained with two independent computer codes.

It should be noted that the Coulomb effects can be screened by the width of the sparticles depending on the specific SUSY scenario. For consistency we will stick to the approach adopted in the NLO calculations, where this screening is not taken into account. In order to study the effects from the hard matching coefficients and the Coulomb corrections separately, we will compare several cross sections with the NLO result and discuss their contribution:
\begin{itemize}
\item The NLL matched cross section is based on the calculations presented in \cite{Beenakker:2009ha,Kulesza:2008jb,Kulesza:2009kq} and will be denoted as $\sigma^{\rm NLO+NLL}$.
\item The NNLL matched cross section without Coulomb contributions to the resummation $\sigma^{\rm NLO+NNLL\; w/o\; Coulomb}$ contains the soft-gluon resummation to NNLL accuracy matched to the full NLO result. The matching is performed according to Eq.~(\ref{eq:matching}). The Coulomb correction to the resummation is not included, so ${\cal C}^{\rm Coul,(1)}$ in Eq.~\eqref{eq:matchingcoeff} is set to zero.
\item The NNLL matched cross section $\sigma^{\rm NLO+NNLL}$ does include the Coulomb contribution ${\cal C}^{\rm Coul,(1)}$ from equation~\eqref{eq:matchingcoeff}. Also in this case Eq.~(\ref{eq:matching}) has been used to match the cross section to the complete NLO result.
\end{itemize}

The NLO cross sections are calculated using the publicly available {\tt PROSPINO} code~\cite{prospino}, based on the calculations presented in Ref.~\cite{Beenakker:1996ch}.  The QCD coupling $\alpha_{\rm s}$ and the parton distribution functions at NLO are defined in the $\overline{\rm MS}$ scheme with five active flavours. The masses of squarks and gluinos are renormalized in the on-shell scheme, and the SUSY particles are decoupled from the running of $\alpha_{\rm s}$ and the parton distribution functions. No top-squark final states are considered.  We sum over squarks with both chiralities ($\tilde{q}_{L}$ and~$\tilde{q}_{R}$), which are taken as mass degenerate. The renormalization and factorization scales $\mu$ are taken to be equal. 

We first discuss the scale dependence of the cross sections. Figure~\ref{fig:scale_dep} shows the squark-antisquark cross section for $m_{\sq}=m_{\gl}=1.2$~TeV as a function of the renormalization and factorization scale $\mu$. The value of $\mu$ is varied around the central scale $\mu_0 = m_\sq$ from $\mu=\mu_0/5$ up to $\mu=5 \,\mu_0$ and results are shown for the LHC at CM energies of 7~TeV~(a) and 14~TeV~(b). 
\FIGURE{
\hspace{-0.55cm}
\begin{tabular}{ll}
(a)\epsfig{file=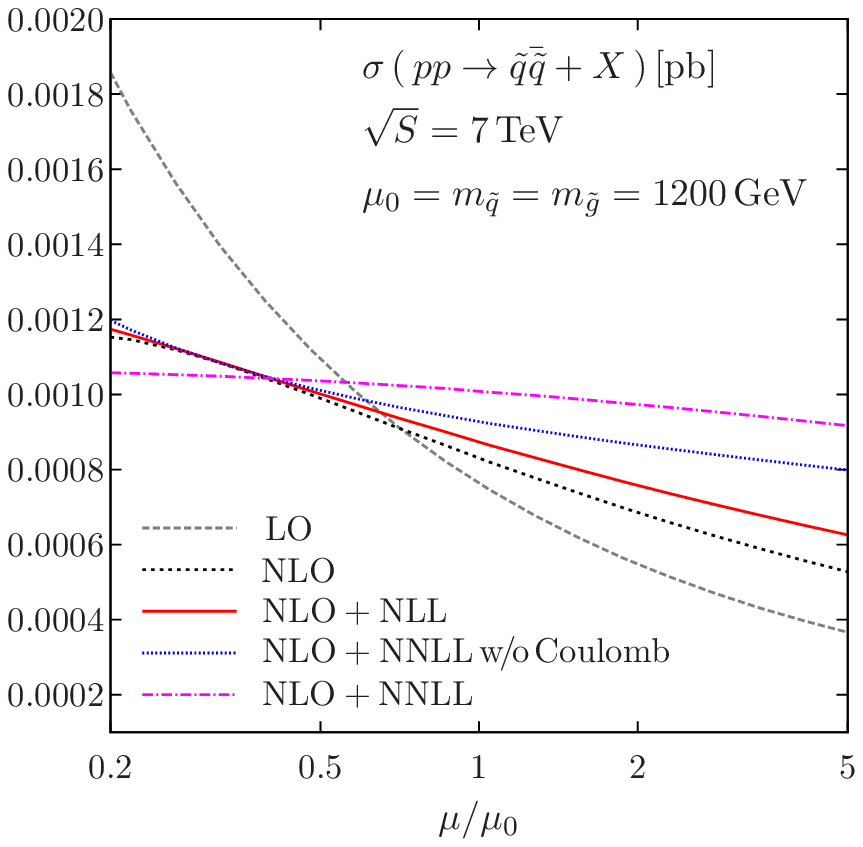, width=0.45\columnwidth}& 
(b)\epsfig{file=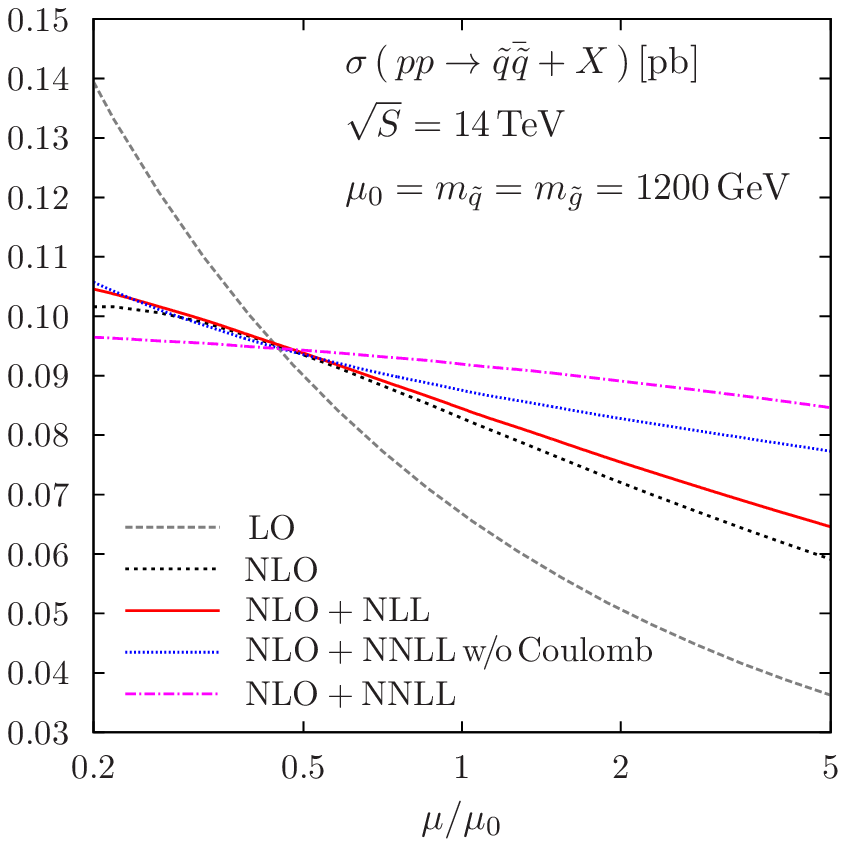, width=0.45\columnwidth}
\end{tabular}
\caption{The scale dependence of the LO, NLO, NLO+NLL and NLO+NNLL (both with the Coulomb part ${\cal C}^{\rm Coul,(1)}$ and without it) squark-antisquark cross sections for the LHC at 7 TeV (a) and 14 TeV (b). The squark and gluino masses have been set to $m_{\sq}=m_{\gl}=1.2$~TeV.\label{fig:scale_dep}}
}

For both collider energies we see the usual scale reduction going from LO to NLO. Including the NLL correction and the NNLL contribution without the Coulomb part ${\cal C}^{\rm Coul,(1)}$ improves the behaviour for moderate values of $\mu/\mu_0$, but a fairly strong scale dependence for small values of $\mu/\mu_0$ remains. Upon inclusion of the Coulomb corrections ${\cal C}^{\rm Coul,(1)}$ the scale dependence stabilises over the whole range.

Figure \ref{fig:scale_unc} shows the mass dependence of the scale uncertainty for the NLO, NLO+NLL and NLO+NNLL cross sections at the LHC. The squark and gluino mass have been taken equal and the scale has been varied in the range $m_\sq/2\le\mu\le 2m_\sq$.
As was to be expected from figure~\ref{fig:scale_dep}, the scale uncertainty reduces as the accuracy of the predictions increases. In the range of squark masses considered here, the NNLL resummation without the Coulomb corrections ${\cal C}^{\rm Coul,(1)}$ already reduces the scale uncertainty to at most 10\% for the LHC at a CM energy of 7~TeV and to even lower values for a CM energy of 14 TeV. The inclusion of the Coulomb term ${\cal C}^{\rm Coul,(1)}$ in the resummed NNLL prediction results in a scale uncertainty of only a few percent for both collider energies. The effect of the threshold resummation is more pronounced for a collider energy of 7~TeV, which is due to the fact that the sparticles are produced closer to threshold in that case.
\FIGURE{
\hspace{-0.55cm}
\begin{tabular}{ll}
(a)\epsfig{file=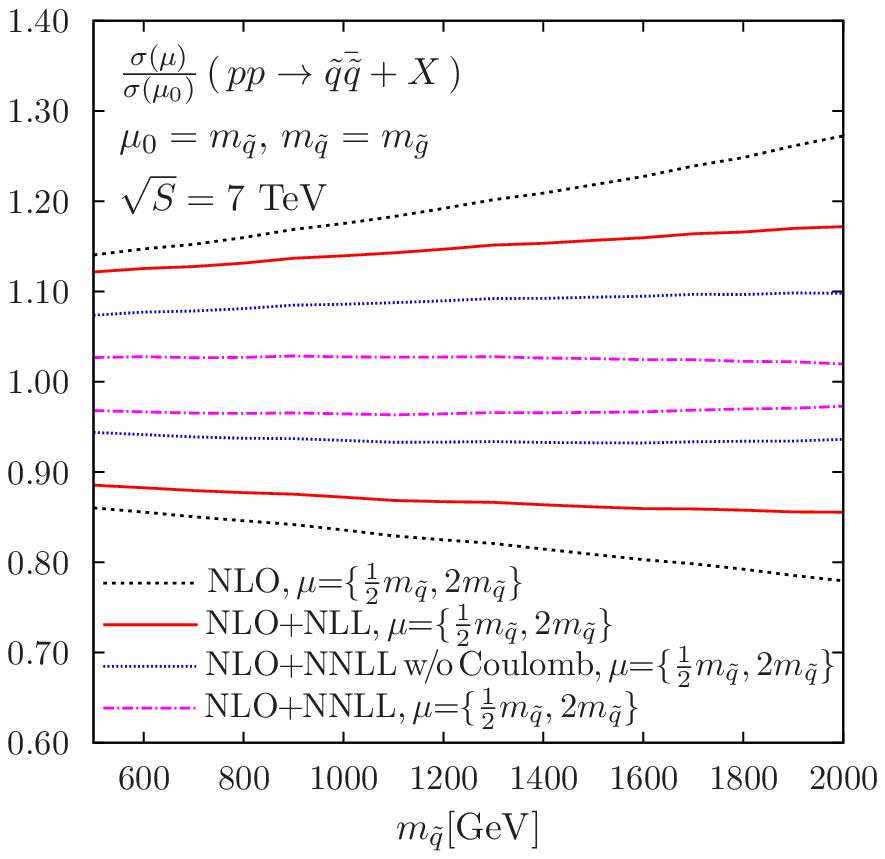, width=0.45\columnwidth}& 
(b)\epsfig{file=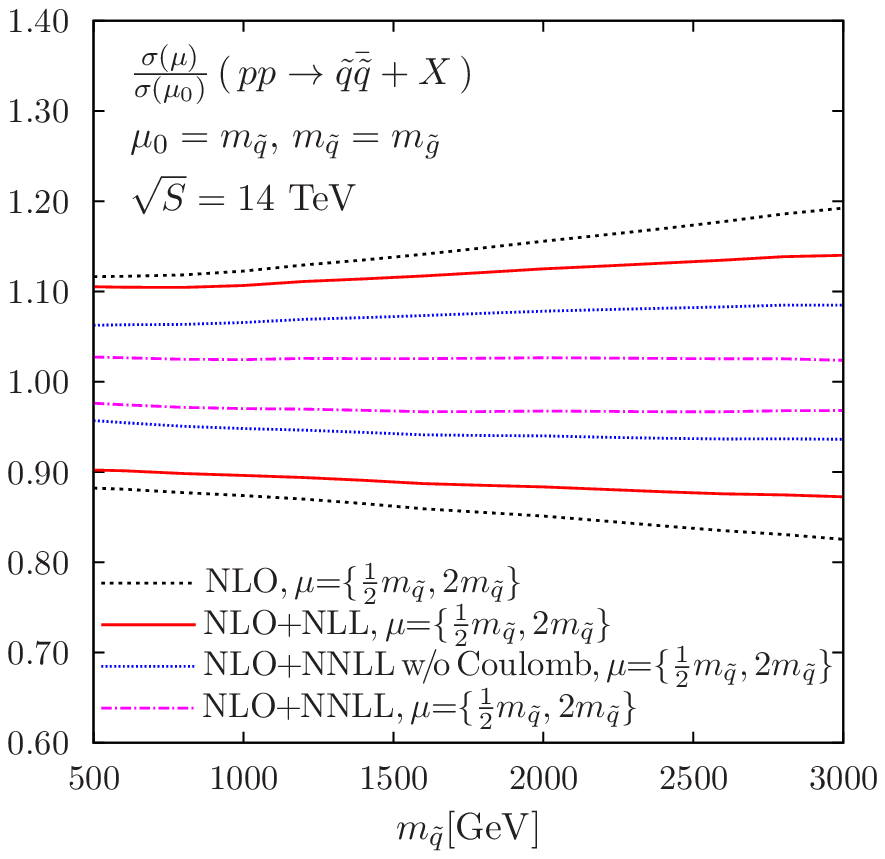, width=0.45\columnwidth}
\end{tabular}
\caption{The scale uncertainty of the NLO, NLO+NLL and NLO+NNLL (both with the Coulomb part ${\cal C}^{\rm Coul,(1)}$ and without it) squark-antisquark cross sections for the LHC at 7 TeV (a) and 14 TeV (b). The common renormalization and factorization scale has been varied in the range $m_\sq/2\le\mu\le 2m_\sq$ and the squark and gluino masses have been taken equal.\label{fig:scale_unc}}
}

Finally we study the $K$-factors with respect to the NLO cross section:
\[K_x=\frac{\sigma^x}{\sigma^{\rm NLO}}~,\]
where $x$ can be NLO+NLL, NLO+NNLL w/o Coulomb or NLO+NNLL. In figure~\ref{fig:K} we study the mass dependence of the $K$-factor for equal squark and gluino masses.
\FIGURE{
\hspace{-0.55cm}
\begin{tabular}{ll}
(a)\epsfig{file=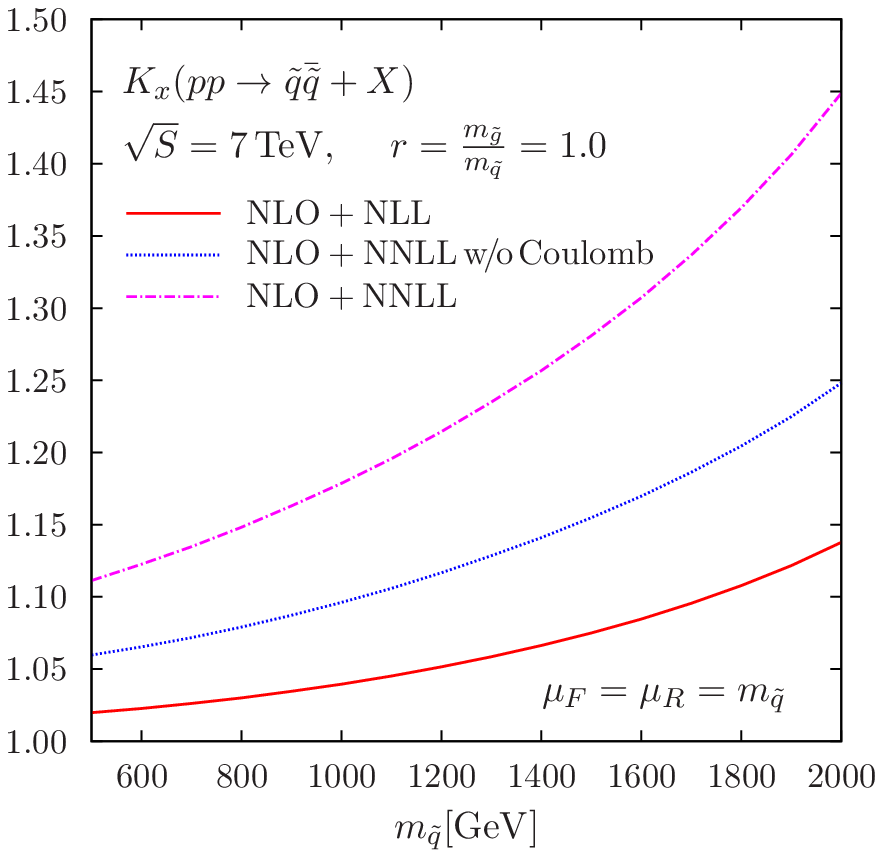, width=0.45\columnwidth}& 
(b)\epsfig{file=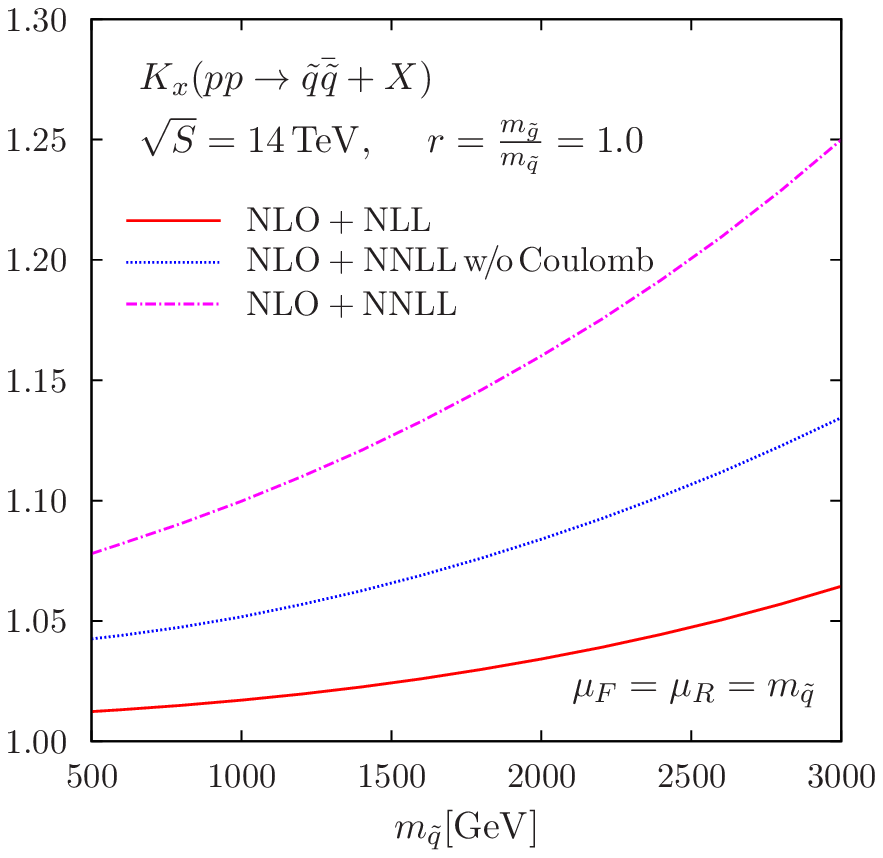, width=0.45\columnwidth}
\end{tabular}\caption{The $K$-factor with respect to the NLO cross section of the NLO+NLL and NLO+NNLL (both with the Coulomb part ${\cal C}^{\rm Coul,(1)}$ and without it) squark-antisquark cross sections for the LHC at 7 TeV (a) and 14 TeV (b). The squark and gluino masses have been taken equal and the common renormalization and factorization scale has been set equal to the squark mass.\label{fig:K}}
}
\FIGURE[h!]{
\begin{tabular}{ccc}
\hspace{1cm} &
\epsfig{file=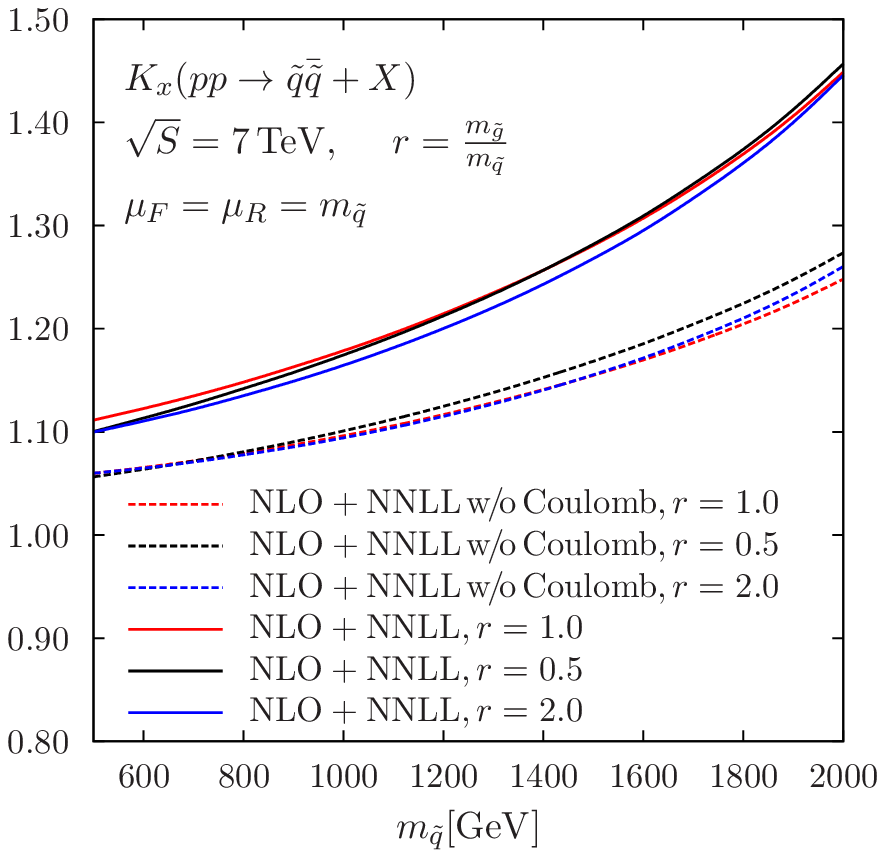, width=0.45\columnwidth} &
\hspace{1cm} 
\end{tabular}
\caption{The $K$-factor with respect to the NLO cross section of the NLO+NNLL squark-antisquark cross sections with and without the Coulomb contributions ${\cal C}^{\rm Coul,(1)}$ for the LHC at 7~TeV. Different ratios of the squark and gluino mass are shown, while the common renormalization and factorization scale has been set equal to the squark mass.\label{fig:K-r}}
}

At the central scale $\mu=m_\sq$ the $K$-factor, and thus the
theoretical prediction of the cross section, increases as more
corrections are included. Also, the effect becomes more pronounced for
higher masses. This was to be expected, since in that case the
particles are produced closer to threshold. As can be seen in
figure~\ref{fig:K}(a), the NNLL resummation without the Coulomb
corrections ${\cal C}^{\rm Coul,(1)}$ already results in a 25\%
increase of the cross section with respect to the NLO cross section
for squarks of 2~TeV and a CM energy of 7~TeV. The contribution from
the Coulomb term to the resummed NLO+NNLL cross section is larger than
the contributions provided by the $g_3$ term in the exponential and
the hard matching coefficient ${\cal C}^{\rm (1)}$, yielding a total
$K$-factor of 1.45. For the case of a CM energy of 14~TeV, which is
shown in figure~\ref{fig:K}(b), the size of the NNLL contributions is
smaller, since the sparticles are produced further away from
threshold. However, for masses of 3~TeV, the $K$-factor for the NNLL
contribution without the Coulomb correction ${\cal C}^{\rm Coul,(1)}$
still yields 1.13, whereas the inclusion of the Coulomb corrections
increases this to 1.25. Although the effect from the Coulomb
corrections could be somewhat smaller in reality due to the finite
lifetime of the squarks, figure~\ref{fig:K} suggests that the NNLL
contribution will remain large.

Figure~\ref{fig:K} only contains the numbers for equal squark and gluino masses, but the effect of the gluino mass is small, as can be seen in figure~\ref{fig:K-r}.
In figure~\ref{fig:K-r} the mass ratio $r=m_\gl/m_\sq$ has been varied. Although some effect can be seen, it is negligible compared to the size of the NNLL corrections. It turns out that this conclusion also holds for a collider energy of 14~TeV. Consequently the NNLL-resummed results are relatively independent of the relation between squark and gluino masses.

The scale dependence of the cross section shows the best stability after including both the hard
matching coefficients ${\cal C}^{(1)}$ and the Coulomb contributions
${\cal C}^{\rm Coul,(1)}$. This indicates that all these contributions
should be taken into account to achieve the observed
cancellation, see also figure~9 in \cite{Beneke:2010da}. However, the observed reduction in the scale dependence
might be modified somewhat by the inclusion of the width of the
particles or by matching to the full NNLO result, which is not
available. In this context we note that, as a consequence of the NNLL
accuracy of resummation, our matched cross section receives additional
non-logarithmic NNLO contributions, which would have been consistently
treated if matching to NNLO had been possible.
A very conservative estimate of the scale uncertainty is provided by the NLO+NNLL w/o Coulomb results, which do not include the Coulomb corrections.\\

\section{Conclusions}
\label{s:conclusion}

We have performed the NNLL resummation of threshold corrections for squark-antisquark hadroproduction. In particular, the previously unknown
hard matching coefficient ${\cal C}^{\rm(1)}$, needed at this level of accuracy, has been calculated analytically. We have also numerically evaluated the NNLL resummed cross section, matched to the NLO fixed-order expression, for squark-antisquark production at the LHC with CM energies of 7 and 14 TeV. At both collider energies the total cross section increases at the central
scale. At 7 TeV collision energy and for a squark mass of 2 TeV, the
NLO+NNLL squark-antisquark cross section is larger than the corresponding
NLO cross section by as much as 45\%. The correction is reduced to 25\% if
the contributions due to Coulombic interactions are not taken into account.
In addition, the scale dependence is reduced significantly, particularly after inclusion of the Coulomb corrections. This information should be used to improve current limits on SUSY masses or, in the case that SUSY is found, to more accurately determine the masses of the sparticles.

\section*{Acknowledgments}

This work has been supported in part by the Helmholtz Alliance
``Physics at the Terascale'', the Foundation for Fundamental
Research of Matter (FOM), program 104 ``Theoretical Particle Physics in
the Era of the LHC", the DFG SFB/TR9 ``Computational Particle Physics'',
and the European Community's Marie-Curie Research Training Network
under contract MRTN-CT-2006-035505 ``Tools and Precision Calculations
for Physics Discoveries at Colliders''. 


\begin{appendix}

\section{The NNLL functions}\label{app:g3}

In the following we list the explicit expression for the NNLL function $g^{(3)}_{ij\to \sq\bar\sq,I}$,
 cf. Refs.~\cite{Moch:2005ba,Moch:2008qy}. Expressions for the LL and NLL functions can be found in Refs.~\cite{Bonciani:1998vc,Kulesza:2009kq}.

The NNLL functions $g^{(3)}_{q\bar q\to \sq\bar\sq,I}$ and $g^{(3)}_{gg\to \sq\bar\sq,I}$ read
\begin{eqnarray}
g^{(3)}_{ij\to \sq\bar\sq,I}\left(\lambda, 4m_{\tilde q}^2,\mu^2 \right) &=& 
\frac{A^{(1)}_i b_1^2}{\pi b_0^4}\frac{1}{1-2\lambda}\left[2\lambda^2+2\lambda \log(1-2\lambda)+\frac{1}{2}\log^2(1-2\lambda)\right]\\ \nn 
&& +\frac{A^{(1)}_i b_2}{\pi b_0^3}\left[2\lambda+\log(1-2\lambda)+\frac{2 \lambda^2}{1-2\lambda}\right]\\ \nn
&& -\frac{2A^{(1)}_i b_1 \gamma_E}{\pi b_0^2}\frac{\left[2\lambda+\log(1-2\lambda)\right]}{1-2\lambda}+\frac{4 A^{(1)}_i}{\pi} \left(\zeta(2)+\gamma_E^2 \right)\frac{\lambda}{1-2\lambda}\\ \nn
&& -\frac{ A^{(2)}_i b_1}{\pi^2 b_0^3}\frac{1}{1-2\lambda}\left[2\lambda(\lambda+1)+\log(1-2\lambda)\right]\\ \nn
&& +\frac{4 A^{(2)}_i \gamma_E}{\pi^2 b_0} \frac{\lambda}{1-2\lambda}+\frac{2 A^{(3)}_i}{\pi^3 b_0^2} \frac{\lambda^2}{1-2\lambda}-\frac{D^{(2)}_i}{\pi^2 b_0}\frac{\lambda}{1-2\lambda} \\ \nn
&& +\frac{D^{(1)}_{ij\to \sq\bar\sq,I}\, b_1}{2 \pi b_0^2 }\frac{\left[2\lambda+\log(1-2\lambda)\right]}{1-2\lambda}-\frac{2 D^{(1)}_{ij\to \sq\bar\sq,I}\,\gamma_E}{\pi}\frac{\lambda}{1-2\lambda}\\ \nn
&& -\frac{D^{(2)}_{ij\to \sq\bar\sq,I}}{\pi^2 b_0}\frac{\lambda}{1-2\lambda}\\ \nn
&& +\left[\frac{A^{(1)}_i b_1}{\pi b_0^2}\frac{2\lambda+\log(1-2\lambda)}{1-2\lambda}-\frac{4 A^{(1)}_i\gamma_E}{\pi}\frac{\lambda}{1-2\lambda}\right] \log\left(\frac{4m_{\tilde q}^2}{\mu^2}\right)\\ \nn
&& +\frac{A^{(1)}_i}{\pi}\frac{\lambda}{1-2\lambda}\log^2\left(\frac{4m_{\tilde q}^2}{\mu^2}\right)-\frac{2 A^{(2)}_i}{\pi^2 b_0}\frac{\lambda}{1-2\lambda}\log\left(\frac{4m_{\tilde q}^2}{\mu^2}\right)\\ \nn
&& +\frac{D^{(1)}_{ij\to \sq\bar\sq,I}}{\pi}\frac{\lambda}{1-2\lambda}\log\left(\frac{4m_{\tilde q}^2}{\mu^2}\right)
\end{eqnarray}
with $\lambda=b_0\alpha_{\rm s}(\mu^2) \log(N) $, $\mu$ the common renormalization and factorization scale, and $\gamma_E$ Euler's constant. The coefficients of the QCD beta function 
are denoted by $b_n$ and the first three coefficients are given by \cite{Tarasov:1980au, Larin:1993tp}
\begin{eqnarray}
b_0&=&\frac{11 C_A-2 n_l}{12 \pi}\,,\\ \nn
b_1&=&\frac{17 C_A^2-5C_A n_l -3C_F  n_l}{24 \pi^2}\,,\\ \nn
b_2&=&\frac{1}{(4 \pi)^3}\left[\frac{2857}{54}C_A^3-\frac{1415}{54}C_A^2 n_l -\frac{205}{18} C_A C_F n_l+C_F^2 n_l 
+\frac{79}{54}C_A n_l^2 +\frac{11}{9}C_F n_l^2  \right]\,,\\ \nn
\end{eqnarray}
where $n_l$ denotes the number of light quark flavours, $C_A=N_c$ and $C_F=\frac{N_c^2-1}{2N_c}$ with $N_c$ the number of colours.
The universal, process independent coefficients up to NNLL accuracy are given by \cite{Moch:2005ba}
\begin{eqnarray}
  A^{(1)}_i &  =  & C_i\,, \\ \nn
  A^{(2)}_i &  =  & \frac{1}{2} C_i \left[ \left( \frac{67}{18} 
     - \zeta(2) \right) C_A - \frac{5}{9} n_l \right]\,, \\ \nn
  A^{(3)}_i & = & \frac{1}{4} C_i \left[ C_A^{2} \left( \frac{245}{24} - \frac{67}{9}\zeta(2) + \frac{11}{6}\zeta(3) 
      + \frac{11}{5}\zeta(2)^{2} \right) 
      + C_F n_l \left( -\frac{55}{24}  + 2\zeta(3)\right) \right. \\ \nn & & \left. \hspace{2cm}
      +\, C_A n_l\, \left( - \frac{209}{108}+ \frac{10}{9}\zeta(2)- \frac{7}{3}\zeta(3) \right) - \frac{n_l^2}{27} \right]\,,
\end{eqnarray}
and \cite{Contopanagos:1996nh,Vogt:2000ci,Catani:2003zt}
\begin{eqnarray}
D^{(2)}_i&=&C_i\left[C_A \left(-\frac{101}{27}+ \frac{11}{3}\zeta(2)+ \frac{7}{2}\zeta(3)\right)+n_l \left(\frac{14}{27}- \frac{2}{3}\zeta(2) \right) \right]\,,
\end{eqnarray}
with the colour factor $C_i=C_F$ for $i=q,\bar{q}$ and $C_i=C_A$ for $i=g$.
The process dependent coefficients read \cite{Bonciani:1998vc}
\begin{eqnarray}
{D^{(1)}_{q\bar q(gg)\to \sq\bar\sq,\bf1}}=0\,, \hspace{1cm } {D^{(1)}_{q\bar q(gg)\to \sq\bar\sq,\bf8(8A,8S)}}=-C_A\,,
\end{eqnarray}
and \cite{Czakon:2009zw, Beneke:2010da}
\begin{eqnarray}
{D^{(2)}_{q\bar q(gg)\to \sq\bar\sq,\bf1}}&=&0\,, \\ \nn
{D^{(2)}_{q\bar q(gg)\to \sq\bar\sq,\bf8(8A,8S)}}&=&-\,C_A\left(\frac{1}{2} \left[ \left( \frac{67}{18} 
     - \zeta(2) \right) C_A - \frac{5}{9} n_l \right]+\frac{C_A}{2}(\zeta(3)-1)+2  \pi b_0 \right)\,.
\end{eqnarray}

\section{Mellin transforms of the Coulomb corrections}\label{app:mellin}

In this appendix we present the analytical results for the Mellin
transforms of the Coulomb corrections in terms of the Euler beta function $\beta$, the digamma function $\Psi$ and the hypergeometric functions $_2F_1$ and $_3F_2$.
For the subprocess $q_{i} \bar q_{j} \to \sq\bar\sq\,$ the expressions for the colour-decomposed Coulomb part in $N$-space are given by
\begin{eqnarray}
\tilde\sigma^{\mathrm{Coul},(1)}_{q_{i} \bar q_{j} \to \sq\bar\sq\,,\bf1}(N) 
&=& -\frac{4\alpha_{\rm s}^3\pi^2}{243 m_{\sq}^2}
    \Bigg[ \frac{4}{N+1}-\frac{4 h}{N+2}\, {_{2}}F_1\left(1,N+2,N+3,h\right) \\ \nn
&& \hspace{1.8cm} +\,4\, I_N(r)+2(r^2-1)\,I_{N+1}(r) \Bigg]\,,\\[3mm]
\tilde\sigma^{\mathrm{Coul},(1)}_{q_{i} \bar q_{j} \to \sq\bar\sq\,,\bf8}(N) 
&=& -\frac{\alpha_{\rm s}^3\pi^2}{2592 m_{\sq}^2}\delta_{ij}
     \Bigg[ \frac{8 n_l}{N^2+3N+2}+\frac{8}{N+1}+(r^2-1)\frac{4}{N+2} \\ \nn
&& \hspace{2.5cm} +\,4\,r^2\, I_{N+1}(r)+(r^2-1)^2\,I_{N+2}(r) \Bigg]\,- \frac{1}{64}\tilde\sigma^{\mathrm{Coul},(1)}_{q_{i} \bar q_{j} \to \sq\bar\sq\,,\bf1}(N)\,,
\end{eqnarray}
whereas for the subprocess $\,gg \to \sq\bar\sq\,\,$ they read
\begin{eqnarray}
\tilde\sigma^{\mathrm{Coul},(1)}_{gg \to \sq\bar\sq\,,\bf1}(N) 
&=& \frac{\alpha_{\rm s}^3\pi^2 n_l}{288 m_{\sq}^2}
    \Bigg[ \frac{2}{N+1}+\frac{2}{N+2}\, +\,2\, I_{N+1}(1)-I_{N+2}(1) \Bigg]\,,\\[3mm]
\tilde\sigma^{\mathrm{Coul},(1)}_{gg \to \sq\bar\sq\,,\bf8A}(N) 
&=& -\frac{\alpha_{\rm s}^3\pi^2 n_l}{1536 m_{\sq}^2}
    \Bigg[ \frac{2}{N+1}+\frac{16}{N+2}\, +\,6\, I_{N+1}(1)+3\,I_{N+2}(1) \Bigg]\,,\\[3mm]
\tilde\sigma^{\mathrm{Coul},(1)}_{gg \to \sq\bar\sq\,,\bf8S}(N)&=&-\frac{5}{16}\tilde\sigma^{\mathrm{Coul},(1)}_{gg \to \sq\bar\sq\,,\bf1}(N)\,.
\end{eqnarray}
Here $r=m_{\tilde g}/m_{\tilde q}$ and $n_l=5$ denotes the number of light quark flavours. 
The function $I_{N}(r)$ is given by 
\begin{eqnarray}
I_{N}(r) &\equiv& \frac{2\sqrt d}{N+1}+2 B_N \log{\left(\frac{r^2+1}{2r}\right)}+B_N\left[\Psi(N+3/2)-\Psi(N+1)\right]\\ \nn 
&& +h\,\frac{N+1}{N+3/2}\, B_N\,\, {_3} F_2(1,1,N+2,2,N+5/2,h)\\ \nn
&&-\frac{d}{2(N+3/2)}  B_N\,\, {_3}F_2(1,1,3/2,2,N+5/2,d)\\ \nn
&& +\frac{2\, d^{3/2}}{3\left(N+1\right)\left(N+2\right)}\,\, {_3}F_2(1,3/2,2,5/2,N+3,d)\\ \nn
&& -\frac{2^{N+2}}{(N+1)^2}\,\, {_3}F_2(-N,N+1,N+1,N+2,N+2,1/2)\,,
\end{eqnarray}
with 
\begin{eqnarray}
h=-\frac{(r^2-1)^2}{4r^2}, \hspace{1.5cm} \mathrm{and} \hspace{1cm} d=\frac{(r^2-1)^2}{(r^2+1)^2}.
\end{eqnarray}
The Euler beta function $\beta (N+1, 1/2)$ is abbreviated by
\begin{eqnarray}
B_{N} &\equiv& \beta (N+1, 1/2).
\end{eqnarray}

\section{How velocity factors can deform the phase space integration}\label{app:vfactors}

As mentioned in section~\ref{s:Ccoeff} the dipole function given in Ref.~\cite{Catani:2002hc} has been modified, which makes it unsuitable for our calculation. We will explicitly show the effect of this modification by considering the change in the second term of Eq.~\eqref{eq:dipole}, which is denoted as $I^{\mathrm{coll}}_{gQ,Q}$ in Eq.~(5.23) of Ref.~\cite{Catani:2002hc}. In Ref.~\cite{Catani:2002hc} finite pieces of the integrand are taken into account as well, but since we showed in section~\ref{s:Ccoeff} that the only contribution at threshold comes from the singular part of the integrand, we will omit these terms.

The singular term of the integrand yields a $1/\epsilon$-pole and a finite piece. The pole cancels the pole of the first term of the dipole function \eqref{eq:dipole}, while the finite piece contributes to the hard matching coefficient. In its unmodified form, the finite piece is given by
\begin{align}
\left.I^{\mathrm{coll,unchanged}}_{gQ,Q}\right|_{\mathrm{fin}}&=2\int_0^{y_+}\mathrm{d}y\left[\frac{1}{y}-\frac{\mu^2\sqrt{[2\mu^2+(1-2\mu^2)(1-y)]^2-4\mu^2}}{y(\mu^2+y(1-2\mu^2))\sqrt{1-4\mu^2}}\right]\nonumber\\
&\approx2\int_0^{2(1-2\mu)}\mathrm{d}y\frac{\sqrt{1-4\mu^2}-\sqrt{(1-y/2)^2-4\mu^2}}{y\sqrt{1-4\mu^2}}\,\label{eq:Icoll}
\end{align}
with
\[y=\frac{p_g\cdot p_j}{p_g\cdot p_j+p_g\cdot p_l+p_j\cdot p_l},\qquad y_+=\frac{1-2\mu}{1-2\mu^2}\quad\mbox{and}\quad\mu=m/\sqrt s\;.\]
The approximation in the second line of Eq.~\eqref{eq:Icoll} is suitable near threshold, where $\mu\approx1/2$. Exactly at threshold the finite part equals $4-4\log(2)$ and exactly cancels the contribution from the first term of Eq.~\eqref{eq:dipole}.

In Ref.~\cite{Catani:2002hc}, the integrand in $I^{\mathrm{coll}}_{gQ,Q}$ has been multiplied by velocity factors in order to simplify the integration:
\[\frac{\tilde v_{gQ,Q}}{v_{gQ,Q}}=\frac{(1-y)\sqrt{1-4\mu^2}}{\sqrt{[2\mu^2+(1-2\mu^2)(1-y)]^2-4\mu^2}}\approx\frac{\sqrt{1-4\mu^2}}{\sqrt{(1-y/2)^2-4\mu^2}}\;,\]
where the approximation in the second step holds near threshold. The velocity factor effectively replaces $\sqrt{1-4\mu^2}$ in the denominator of Eq.~\eqref{eq:Icoll} by $\sqrt{(1-y/2)^2-4\mu^2}$, which amounts to a shift comparable in size to the value of the numerator. In the strict soft limit $y$ vanishes and the velocity factors have no effect. However, we are integrating over gluons that are not soft compared to the energy above threshold $\sqrt{s}-2m$, so we also need the correct behaviour away from the strict soft limit. In fact, if the velocity factors are included the integral $I^{\mathrm{coll}}_{gQ,Q}$ vanishes at threshold, so it does no longer cancel the contribution from the first term of Eq.~\eqref{eq:dipole}.

Usually the velocity factors do not pose a problem in calculations using dipole subtraction, since the terms are subtracted from the real part and added to the virtual part. Therefore it does not matter if a dipole function is deformed, as long as the pole is reproduced. Finite contributions can always be moved between $\sigma^{\{2\}}$ and $\sigma^{\{3\}}$. However, we argued that $\sigma^{\{3\}}$ vanishes at threshold due to phase-space suppression, which is not true if the phase space integration is deformed by velocity factors. Therefore we need the unchanged dipole function for this particular calculation.

\section{The hard matching coefficients for squark-antisquark production}\label{app:Ccoeff}

Here we present the exact expressions for the hard matching coefficients ${\cal C}^{\rm(1)}$ for the squark-antisquark production processes as defined in Eq.~\eqref{eq:matchingcoeff}. We sum over squarks with both chiralities ($\tilde{q}_{L}$ and $\tilde{q}_{R}$). No top-squark final states are considered and all squarks are considered to be mass-degenerate with mass $m_\sq$. Top squarks are taken into account in the loops, where they are taken to be mass-degenerate with the other squarks. The calculation is outlined in section~\ref{s:Ccoeff} and was done with FORM \cite{Vermaseren:2000nd}. We first define:
\[\beta_{12}(q^2)=\sqrt{1 - \frac{4m_1m_2}{q^2 - (m_1 - m_2)^2}}~,\qquad x_{12}(q^2)=\frac{\beta_{12}(q^2)-1}{\beta_{12}(q^2)+1}\]
and
\[m_-^2=m_{\tilde g}^2-m_{\tilde q}^2,\qquad m_+^2=m_{\tilde g}^2+m_{\tilde q}^2,\]
where $m_\gl$ is the gluino mass. Denoting the number of light flavours by $n_l=5$, the total number of flavours by $n_f=6$ and the number of colours by $N_c=3$, we also define:
\begin{align*}
\gamma_q&=\frac{3}{2}C_F&C_F&=\frac{N_c^2-1}{2N_c}\\
\gamma_g&=\frac{11}{6}C_A-\frac{1}{3}n_l&C_A&=N_c\,.
\end{align*}
We denote the factorization scale by $\mu_F$, the renormalization scale by $\mu_R$ and Euler's constant by $\gamma_E$.
For the $q\bar q\to\tilde q\bar{\tilde q}$ process both the singlet $\bf1$ and the octet $\bf8$ representation contribute:
\begin{align*}
{\cal C}_{q\bar q\to\tilde q\bar{\tilde q},I}^{(1)}&=\mathrm{Re}\Bigg\{\frac{2C_F}{3}\pi^2+\gamma_g\log\bigg(\frac{\mu_R^2}{m_{\tilde q}^2}\bigg)-\gamma_q\log\bigg(\frac{\mu_F^2}{m_{\tilde q}^2}\bigg)+F_0(m_{\tilde q},m_{\tilde g},m_t)+\frac{19N_c}{24}\\
&\quad+\frac{23}{8N_c}+\frac{1-3N_c^2}{N_c}\log\bigg(\frac{m_+^2}{m_{\tilde q}^2}\bigg)-\frac{2}{N_c}\log\left(2\right)+\bigg(\frac{7N_c}{6}+\frac{2m_{\tilde g}^2}{m_+^2}C_F\bigg)\log\bigg(\frac{m_{\tilde g}^2}{m_{\tilde q}^2}\bigg)\\
&\quad-\frac{m_{\tilde g}^2}{m_{\tilde q}^2}\bigg(\frac{m_-^2}{m_{\tilde q}^2}\log\bigg(\frac{m_-^2}{m_{\tilde g}^2}\bigg)+1\bigg)C_F+\frac{m_{\tilde g}^2}{2m_-^2}\bigg(\frac{m_{\tilde g}^2}{m_-^2}\log\bigg(\frac{m_{\tilde g}^2}{m_{\tilde q}^2}\bigg)-1\bigg)C_F\\
&\quad-\frac{1}{2N_c}\bigg(\frac{m_{\tilde g}^2}{m_{\tilde q}^2}-3\bigg)F_1\left(m_{\tilde q},m_{\tilde g}\right)+\bigg(\frac{m_+^2}{2m_{\tilde q}^2}C_F+\frac{1}{N_c}\bigg)F_2\left(m_{\tilde q},m_{\tilde g}\right)\\
&\quad+2C_F\bigg(\gamma_E^2 - 2\gamma_E\log(2) + \gamma_E\log\bigg(\frac{\mu_F^2}{m_\sq^2}\bigg)\bigg)\\
&\quad+\bigg[-\frac{\pi^2}{4}+\log\bigg(\frac{m_+^2}{m_{\tilde q}^2}\bigg)-\log\left(2\right)-\frac{m_{\tilde g}^2}{m_+^2}\log\bigg(\frac{m_{\tilde g}^2}{m_{\tilde q}^2}\bigg)+2+\gamma_E\\
&\quad-\frac{1}{4}\bigg(\frac{m_{\tilde g}^2}{m_{\tilde q}^2}-3\bigg)\left(F_1\left(m_{\tilde q},m_{\tilde g}\right)+F_2\left(m_{\tilde q},m_{\tilde g}\right)\right)\bigg]C_2(I)\Bigg\}.
\end{align*}
In this equation the last two lines are proportional to the quadratic Casimir invariant of the representation, which is zero for the singlet and $N_c$ for the octet representation. We have defined the functions:
\begin{align*}
F_0(m_{\tilde q},m_{\tilde g},m_t)&=\frac{m_t^2}{2m_{\tilde g}^2}-\bigg(1+\frac{m_{\tilde q}^2}{2m_{\tilde g}^2}\bigg)n_f+\bigg(\frac{m_-^6}{2m_+^2m_{\tilde g}^4}\log\bigg(\frac{m_-^2}{m_{\tilde q}^2}\bigg)+\frac{4m_{\tilde q}^2}{m_+^2}\log\left(2\right)\!\bigg)n_l\\
&\quad+\bigg(\frac{m_t^4}{2m_{\tilde q}^2m_{\tilde g}^2}-\frac{(m_{\tilde q}^2-m_t^2)^2}{4m_{\tilde g}^4}+\frac{m_{\tilde q}^2-m_t^2}{m_{\tilde g}^2}-\frac{1}{12}\bigg)\log\bigg(\frac{m_t^2}{m_{\tilde q}^2}\bigg)\\
&\quad-\frac{m_-^2\big(m_{\tilde g}^2-(m_{\tilde q}-m_t)^2\big)\big(m_{\tilde g}^2-m_{\tilde q}^2+m_t^2\big)}{2m_{\tilde g}^4m_+^2}\beta_{\tilde qt}(m_{\tilde g}^2)\log\left(x_{\tilde qt}(m_{\tilde g}^2)\right)\\
&\quad+\frac{m_t^4-2m_{\tilde q}m_t^3+4m_{\tilde q}^3m_t-4m_{\tilde q}^4}{m_{\tilde q}^2m_+^2}\beta_{\tilde qt}(-m_{\tilde q}^2)\log\left(x_{\tilde qt}(-m_{\tilde q}^2)\right)\\
F_1(m_{\tilde q},m_{\tilde g})&=\mathrm{Li}_2\bigg(\frac{m_-^2}{2m_{\tilde g}^2}\bigg)+\mathrm{Li}_2\bigg(1-\frac{m_-^2}{2m_{\tilde q}^2}\bigg)+\frac{\pi^2}{12}+\log\bigg(\frac{m_-^2}{2m_{\tilde g}^2}\bigg)\log\bigg(\frac{m_+^2}{2m_{\tilde q}^2}\bigg)\\
&\quad+\frac{1}{2}\log^2\!\bigg(\frac{m_{\tilde g}^2}{m_{\tilde q}^2}\bigg)\\
F_2(m_{\tilde q},m_{\tilde g})&=\mathrm{Li}_2\bigg(\frac{m_{\tilde q}^2}{m_{\tilde g}^2}\bigg)-\mathrm{Li}_2\bigg(-\frac{m_{\tilde q}^2}{m_{\tilde g}^2}\bigg)+\log\bigg(\frac{m_+^2}{m_-^2}\bigg)\log\bigg(\frac{m_{\tilde g}^2}{m_{\tilde q}^2}\bigg)\,.
\end{align*}
For the $gg\to\tilde q\bar{\tilde q}$ process the antisymmetric octet $\bf8_A$ vanishes because it yields a $p$-wave contribution, which vanishes at threshold. The hard matching coefficients for the singlet $\bf1$ and the symmetric octet $\bf8_S$ do contribute:
\begin{align*}
{\cal C}_{gg\to\tilde q\bar{\tilde q},\bf8_A}^{(1)}&=0\\
{\cal C}_{gg\to\tilde q\bar{\tilde q},I}^{(1)}&=\mathrm{Re}\Bigg\{\pi^2\bigg(\frac{5N_c}{12}-\frac{C_F}{4}\bigg)+\gamma_g\log\bigg(\frac{\mu_R^2}{\mu_F^2}\bigg)-\frac{m_{\tilde g}^2N_c}{2m_{\tilde q}^2}\log^2\left(x_{\tilde g\tilde g}(4m_{\tilde q}^2)\right)\\
&\quad+C_F\bigg(\frac{m_+^2m_-^2}{2m_{\tilde q}^4}\log\bigg(\frac{m_+^2}{m_-^2}\bigg)-\frac{m_{\tilde g}^2}{m_{\tilde q}^2}-3\bigg)+\frac{m_+^2N_c}{2m_{\tilde q}^2}\bigg(\mathrm{Li}_2\bigg(\!-\!\frac{m_{\tilde q}^2}{m_{\tilde g}^2}\bigg)-\mathrm{Li}_2\bigg(\frac{m_{\tilde q}^2}{m_{\tilde g}^2}\bigg)\bigg)\\
&\quad+2C_A\bigg(\gamma_E^2 - 2\gamma_E\log(2) + \gamma_E\log\bigg(\frac{\mu_F^2}{m_\sq^2}\bigg)\bigg)\\
&\quad+\bigg[\frac{\pi^2}{8}-\frac{1}{2}\mathrm{Li}_2\bigg(\!\!-\!\frac{m_{\tilde q}^2}{m_{\tilde g}^2}\bigg)+\frac{1}{2}\mathrm{Li}_2\bigg(\frac{m_{\tilde q}^2}{m_{\tilde g}^2}\bigg)+\frac{m_{\tilde g}^2}{4m_{\tilde q}^2}\log^2\!\left(x_{\tilde g\tilde g}(4m_{\tilde q}^2)\right)+2+\gamma_E\bigg]C_2(I)\Bigg\}
\end{align*}
where in the second equation the representation $I$ can be the $\bf1$ or the $\bf8_S$ and the last line is proportional to the quadratic Casimir invariant of the representation.

\end{appendix}

\bibliographystyle{JHEP}

\begin{thebibliography}{10}

\bibitem{Golfand:1971iw}
Y.~Golfand and E.~Likhtman, {\it {Extension of the Algebra of Poincare Group
  Generators and Violation of p Invariance}},  {\em JETP Lett.} {\bf 13} (1971)
  323--326. See also
  \url{http://www.jetpletters.ac.ru/ps/717/article_11110.shtml} Russian
  version.

\bibitem{Wess:1974tw}
J.~Wess and B.~Zumino, {\it {Supergauge Transformations in Four-Dimensions}},
  {\em Nucl.Phys.} {\bf B70} (1974) 39--50.

\bibitem{Nilles:1983ge}
H.~P. Nilles, {\it {Supersymmetry, Supergravity and Particle Physics}},  {\em
  Phys. Rept.} {\bf 110} (1984) 1--162.

\bibitem{Haber:1984rc}
H.~E. Haber and G.~L. Kane, {\it {The Search for Supersymmetry: Probing Physics
  Beyond the Standard Model}},  {\em Phys. Rept.} {\bf 117} (1985) 75--263.



\bibitem{Aad:2011ib}
  {\bf ATLAS Collaboration}, G.~Aad {\em et.~al.},
{\it Search for squarks and gluinos using final states with jets and missing transverse momentum with the ATLAS detector in sqrt(s) = 7 TeV proton-proton collisions},
  [\href{http://xxx.lanl.gov/abs/1109.6572}{{\tt arXiv:1109.6572}}].

\bibitem{Chatrchyan:2011zy}
  {\bf CMS Collaboration}, S.~Chatrchyan {\em et.~al.},
  {\it Search for Supersymmetry at the LHC in Events with Jets and Missing Transverse Energy},
  [\href{http://xxx.lanl.gov/abs/1109.2352}{{\tt arXiv:1109.2352}}].

\bibitem{Aad:2009wy}
{\bf ATLAS Collaboration}, G.~Aad {\em et.~al.}, {\it
  {Expected Performance of the ATLAS Experiment - Detector, Trigger and
  Physics}},  tech. rep., 2009.

\bibitem{Bayatian:2006zz}
{\bf CMS Collaboration}, G.~Bayatian {\em et.~al.}, {\it {CMS
  technical design report, volume II: Physics performance}},  {\em J.Phys.G}
  {\bf G34} (2007) 995--1579.

\bibitem{Baer:2007ya}
H.~Baer, V.~Barger, G.~Shaughnessy, H.~Summy, and L.-t. Wang, {\it {Precision
  gluino mass at the LHC in SUSY models with decoupled scalars}},  {\em
  Phys.Rev.} {\bf D75} (2007) 095010,
  [\href{http://xxx.lanl.gov/abs/hep-ph/0703289}{{\tt hep-ph/0703289}}].

\bibitem{Dreiner:2010gv}
H.~K. Dreiner, M.~Kramer, J.~M. Lindert, and B.~O'Leary, {\it {SUSY parameter
  determination at the LHC using cross sections and kinematic edges}},  {\em
  JHEP} {\bf 1004} (2010) 109, [\href{http://xxx.lanl.gov/abs/1003.2648}{{\tt
  arXiv:1003.2648}}].

\bibitem{Kane:1982hw}
G.~L. Kane and J.~Leveille, {\it {Experimental Constraints on Gluino Masses and
  Supersymmetric Theories}},  {\em Phys.Lett.} {\bf B112} (1982) 227.

\bibitem{Dawson:1983fw}
S.~Dawson, E.~Eichten, and C.~Quigg, {\it {Search for Supersymmetric Particles
  in Hadron - Hadron Collisions}},  {\em Phys.Rev.} {\bf D31} (1985) 1581.

\bibitem{Beenakker:1994an}
W.~Beenakker, R.~Hopker, M.~Spira, and P.~Zerwas, {\it {Squark production at
  the Tevatron}},  {\em Phys.Rev.Lett.} {\bf 74} (1995) 2905--2908,
  [\href{http://xxx.lanl.gov/abs/hep-ph/9412272}{{\tt hep-ph/9412272}}].

\bibitem{Beenakker:1996ch}
W.~Beenakker, R.~Hopker, M.~Spira, and P.~M. Zerwas, {\it {Squark and gluino
  production at hadron colliders}},  {\em Nucl. Phys.} {\bf B492} (1997)
  51--103, [\href{http://xxx.lanl.gov/abs/hep-ph/9610490}{{\tt
  hep-ph/9610490}}].

\bibitem{Hollik:2008yi}
W.~Hollik and E.~Mirabella, {\it {Squark anti-squark pair production at the
  LHC: The Electroweak contribution}},  {\em JHEP} {\bf 0812} (2008) 087,
  [\href{http://xxx.lanl.gov/abs/0806.1433}{{\tt arXiv:0806.1433}}].

\bibitem{Bornhauser:2007bf}
S.~Bornhauser, M.~Drees, H.~K. Dreiner, and J.~S. Kim, {\it {Electroweak
  contributions to squark pair production at the LHC}},  {\em Phys.Rev.} {\bf
  D76} (2007) 095020, [\href{http://xxx.lanl.gov/abs/0709.2544}{{\tt
  arXiv:0709.2544}}].

\bibitem{Sterman:1986aj}
G.~F. Sterman, {\it {Summation of Large Corrections to Short Distance Hadronic
  Cross-Sections}},  {\em Nucl.Phys.} {\bf B281} (1987) 310.

\bibitem{Catani:1989ne}
S.~Catani and L.~Trentadue, {\it {Resummation of the QCD Perturbative Series
  for Hard Processes}},  {\em Nucl. Phys.} {\bf B327} (1989) 323.

\bibitem{BBPC:BBPC19400460520}
A.~Sommerfeld, {\em Atombau und Spektrallinien, II. Band}.
\newblock Braunschweig: Vieweg, 1939, p. 138.

\bibitem{Sakharov:1948yq}
A.~Sakharov, {\it {Interaction of an electron and positron in pair
  production}},  {\em Zh.Eksp.Teor.Fiz.} {\bf 18} (1948) 631--635.

\bibitem{Catani:1996dj}
S.~Catani, M.~L. Mangano, P.~Nason, and L.~Trentadue, {\it {The Top
  cross-section in hadronic collisions}},  {\em Phys.Lett.} {\bf B378} (1996)
  329--336, [\href{http://xxx.lanl.gov/abs/hep-ph/9602208}{{\tt
  hep-ph/9602208}}].

\bibitem{Fadin:1990wx}
V.~S. Fadin, V.~A. Khoze, and T.~Sjostrand, {\it {On the threshold behavior of
  heavy top production}},  {\em Z.Phys.} {\bf C48} (1990) 613--622.

\bibitem{Kiyo:2008bv}
Y.~Kiyo, J.~H. Kuhn, S.~Moch, M.~Steinhauser, and P.~Uwer, {\it {Top-quark pair
  production near threshold at LHC}},  {\em Eur.Phys.J.} {\bf C60} (2009)
  375--386, [\href{http://xxx.lanl.gov/abs/0812.0919}{{\tt arXiv:0812.0919}}].

\bibitem{Hagiwara:2008df}
K.~Hagiwara, Y.~Sumino, and H.~Yokoya, {\it {Bound-state Effects on Top Quark
  Production at Hadron Colliders}},  {\em Phys.Lett.} {\bf B666} (2008) 71--76,
  [\href{http://xxx.lanl.gov/abs/0804.1014}{{\tt arXiv:0804.1014}}].

\bibitem{Hagiwara:2009hq}
  K.~Hagiwara and H.~Yokoya,
  {\it {Bound-state effects on gluino-pair production at hadron colliders}},
  {\em JHEP} {\bf 0910} (2009) 049,
  [\href{http://xxx.lanl.gov/abs/0909.3204}{{\tt  arXiv:0909.3204}}].


\bibitem{Kauth:2011bz}
M.~R. Kauth, A.~Kress, and J.~H. Kuhn, {\it {Gluino-Squark Production at the
  LHC: The Threshold}},  \href{http://xxx.lanl.gov/abs/1108.0542}{{\tt
  arXiv:1108.0542}}.

\bibitem{Kauth:2011vg}
M.~R. Kauth, J.~H. Kuhn, P.~Marquard, and M.~Steinhauser, {\it {Gluino Pair
  Production at the LHC: The Threshold}},
  \href{http://xxx.lanl.gov/abs/1108.0361}{{\tt arXiv:1108.0361}}.

\bibitem{Beneke:2009rj}
M.~Beneke, P.~Falgari, and C.~Schwinn, {\it {Soft radiation in heavy-particle
  pair production: All-order colour structure and two-loop anomalous
  dimension}},  {\em Nucl.Phys.} {\bf B828} (2010) 69--101,
  [\href{http://xxx.lanl.gov/abs/0907.1443}{{\tt arXiv:0907.1443}}].

\bibitem{Beneke:2010da}
M.~Beneke, P.~Falgari, and C.~Schwinn, {\it {Threshold resummation for pair
  production of coloured heavy (s)particles at hadron colliders}},  {\em
  Nucl.Phys.} {\bf B842} (2011) 414--474,
  [\href{http://xxx.lanl.gov/abs/1007.5414}{{\tt arXiv:1007.5414}}].

\bibitem{Kulesza:2008jb}
A.~Kulesza and L.~Motyka, {\it {Threshold resummation for squark-antisquark and
  gluino- pair production at the LHC}},  {\em Phys. Rev. Lett.} {\bf 102}
  (2009) 111802, [\href{http://xxx.lanl.gov/abs/0807.2405}{{\tt
  arXiv:0807.2405}}].

\bibitem{Kulesza:2009kq}
A.~Kulesza and L.~Motyka, {\it {Soft gluon resummation for the production of
  gluino-gluino and squark-antisquark pairs at the LHC}},  {\em Phys. Rev.}
  {\bf D80} (2009) 095004, [\href{http://xxx.lanl.gov/abs/0905.4749}{{\tt
  arXiv:0905.4749}}].

\bibitem{Beenakker:2009ha}
W.~Beenakker, S.~Brensing, M.~Kr\"amer, A.~Kulesza, E.~Laenen, and I.~Niessen,
  {\it {Soft-gluon resummation for squark and gluino hadroproduction}},  {\em
  JHEP} {\bf 12} (2009) 41, [\href{http://xxx.lanl.gov/abs/0909.4418}{{\tt
  arXiv:0909.4418}}].

\bibitem{Beenakker:2010nq}
W.~Beenakker, S.~Brensing, M.~Kr\"amer, A.~Kulesza, E.~Laenen, and I.~Niessen,
  {\it {Supersymmetric top and bottom squark production at hadron colliders}},
  {\em JHEP} {\bf 8} (2010) 1--25,
  [\href{http://xxx.lanl.gov/abs/1006.4771}{{\tt arXiv:1006.4771}}].

\bibitem{Beenakker:2011fu}
W.~Beenakker, S.~Brensing, M.~Kr\"amer, A.~Kulesza, E.~Laenen, L.~Motyka, and
  I.~Niessen, {\it {Squark and gluino hadroproduction}},  {\em Int.J.Mod.Phys.}
  {\bf A26} (2011) 2637--2664, [\href{http://xxx.lanl.gov/abs/1105.1110}{{\tt
  arXiv:1105.1110}}].

\bibitem{Langenfeld:2009eg}
U.~Langenfeld and S.-O. Moch, {\it {Higher-order soft corrections to squark
  hadro-production}},  {\em Phys.Lett.} {\bf B675} (2009) 210--221,
  [\href{http://xxx.lanl.gov/abs/0901.0802}{{\tt arXiv:0901.0802}}].



\bibitem{Bonciani:1998vc}
R.~Bonciani, S.~Catani, M.~L. Mangano, and P.~Nason, {\it {NLL resummation of
  the heavy-quark hadroproduction cross- section}},  {\em Nucl. Phys.} {\bf
  B529} (1998) 424--450, [\href{http://xxx.lanl.gov/abs/hep-ph/9801375}{{\tt
  hep-ph/9801375}}].

\bibitem{Kidonakis:2001nj}
N.~Kidonakis, E.~Laenen, S.~Moch, and R.~Vogt, {\it {Sudakov resummation and
  finite order expansions of heavy quark hadroproduction cross sections}},
  {\em Phys. Rev.} {\bf D64} (2001) 114001,
  [\href{http://xxx.lanl.gov/abs/hep-ph/0105041}{{\tt
      hep-ph/0105041}}].

\bibitem{Kidonakis:2010dk}
  N.~Kidonakis,
  {\it {Next-to-next-to-leading soft-gluon corrections for the top quark cross section and transverse momentum distribution}},
  {\em Phys. Rev.}  {\bf D82 } (2010)  114030,
  [\href{http://xxx.lanl.gov/abs/1009.4935}{{\tt
      arXiv:1009.4935}}]. 

\bibitem{Ahrens:2010zv}
  V.~Ahrens, A.~Ferroglia, M.~Neubert, B.~D.~Pecjak, L.~L.~Yang,
  {\it {Renormalization-Group Improved Predictions for Top-Quark Pair Production at Hadron Colliders}},
  {\em JHEP} {\bf 1009 } (2010)  097,   [\href{http://xxx.lanl.gov/abs/1003.5827}{{\tt
      arXiv:1003.5827}}]. 


\bibitem{Beneke:2011mq}
  M.~Beneke, P.~Falgari, S.~Klein, C.~Schwinn,
  {\it {Hadronic top-quark pair production with NNLL threshold resummation}},
   [\href{http://xxx.lanl.gov/abs/1109.1536}{{\tt
      arXiv:1109.1536}}].

\bibitem{Ellis:1983ed}
J.~R. Ellis and S.~Rudaz, {\it {Search for Supersymmetry in Toponium Decays}},
  {\em Phys.Lett.} {\bf B128} (1983) 248.

\bibitem{Contopanagos:1996nh}
H.~Contopanagos, E.~Laenen, and G.~Sterman, {\it {Sudakov factorization and
  resummation}},  {\em Nucl. Phys.} {\bf B484} (1997) 303--330,
  [\href{http://xxx.lanl.gov/abs/hep-ph/9604313}{{\tt hep-ph/9604313}}].

\bibitem{Kidonakis:1998bk}
N.~Kidonakis, G.~Oderda, and G.~Sterman, {\it {Threshold resummation for dijet
  cross sections}},  {\em Nucl. Phys.} {\bf B525} (1998) 299--332,
  [\href{http://xxx.lanl.gov/abs/hep-ph/9801268}{{\tt hep-ph/9801268}}].

\bibitem{Kidonakis:1998nf}
N.~Kidonakis, G.~Oderda, and G.~Sterman, {\it {Evolution of color exchange in
  {QCD} hard scattering}},  {\em Nucl. Phys.} {\bf B531} (1998) 365--402,
  [\href{http://xxx.lanl.gov/abs/hep-ph/9803241}{{\tt hep-ph/9803241}}].

\bibitem{Botts:1989kf}
J.~Botts and G.~Sterman, {\it {Hard Elastic Scattering in QCD: Leading
  Behavior}},  {\em Nucl. Phys.} {\bf B325} (1989) 62.

\bibitem{Kidonakis:1997gm}
N.~Kidonakis and G.~Sterman, {\it {Resummation for QCD hard scattering}},  {\em
  Nucl. Phys.} {\bf B505} (1997) 321--348,
  [\href{http://xxx.lanl.gov/abs/hep-ph/9705234}{{\tt hep-ph/9705234}}].

\bibitem{Moch:2005ba}
S.~Moch, J.~Vermaseren, and A.~Vogt, {\it {Higher-order corrections in
  threshold resummation}},  {\em Nucl.Phys.} {\bf B726} (2005) 317--335,
  [\href{http://xxx.lanl.gov/abs/hep-ph/0506288}{{\tt hep-ph/0506288}}].

\bibitem{Czakon:2009zw}
M.~Czakon, A.~Mitov, and G.~Sterman, {\it {Threshold Resummation for Top-Pair
  Hadroproduction to Next-to-Next-to-Leading Log}},  {\em Phys. Rev.} {\bf D80}
  (2009) 074017, [\href{http://xxx.lanl.gov/abs/0907.1790}{{\tt
  arXiv:0907.1790}}].

\bibitem{Moch:2008qy}
S.~Moch and P.~Uwer, {\it {Theoretical status and prospects for top-quark pair
  production at hadron colliders}},  {\em Phys.Rev.} {\bf D78} (2008) 034003,
  [\href{http://xxx.lanl.gov/abs/0804.1476}{{\tt arXiv:0804.1476}}].

\bibitem{Catani:1996yz}
S.~Catani, M.~L. Mangano, P.~Nason, and L.~Trentadue, {\it {The Resummation of
  soft gluons in hadronic collisions}},  {\em Nucl.Phys.} {\bf B478} (1996)
  273--310, [\href{http://xxx.lanl.gov/abs/hep-ph/9604351}{{\tt
  hep-ph/9604351}}].

\bibitem{Kulesza:2003wn}
A.~Kulesza, G.~F. Sterman, and W.~Vogelsang, {\it {Joint resummation for Higgs
  production}},  {\em Phys.Rev.} {\bf D69} (2004) 014012,
  [\href{http://xxx.lanl.gov/abs/hep-ph/0309264}{{\tt hep-ph/0309264}}].

\bibitem{Czakon:2008cx}
M.~Czakon and A.~Mitov, {\it {On the Soft-Gluon Resummation in Top Quark Pair
  Production at Hadron Colliders}},  {\em Phys. Lett.} {\bf B680} (2009)
  154--158, [\href{http://xxx.lanl.gov/abs/0812.0353}{{\tt arXiv:0812.0353}}].

\bibitem{Catani:1996vz}
S.~Catani and M.~H. Seymour, {\it {A general algorithm for calculating jet
  cross sections in NLO QCD}},  {\em Nucl. Phys.} {\bf B485} (1997) 291--419,
  [\href{http://xxx.lanl.gov/abs/hep-ph/9605323}{{\tt hep-ph/9605323}}].

\bibitem{Catani:2002hc}
S.~Catani, S.~Dittmaier, M.~H. Seymour, and Z.~Trocsanyi, {\it {The dipole
  formalism for next-to-leading order QCD calculations with massive partons}},
  {\em Nucl. Phys.} {\bf B627} (2002) 189--265,
  [\href{http://xxx.lanl.gov/abs/hep-ph/0201036}{{\tt hep-ph/0201036}}].

\bibitem{Martin:2009iq}
A.~D. Martin, W.~J. Stirling, R.~S. Thorne, and G.~Watt, {\it {Parton
  distributions for the LHC}},  {\em Eur. Phys. J.} {\bf C63} (2009) 189--285,
  [\href{http://xxx.lanl.gov/abs/0901.0002}{{\tt arXiv:0901.0002}}].

\bibitem{PDG}
K. Nakamura et al. (Particle Data Group), {\em J. Phys. G} {\bf 37},
(2010) 075021 
and 2011 partial update for the 2012 edition. 

\bibitem{prospino}
see \url{http://www.thphys.uni-heidelberg.de/~plehn/prospino/} or
  \url{http://people.web.psi.ch/spira/prospino/}.

\bibitem{Vermaseren:2000nd}
J.~Vermaseren, {\it {New features of FORM}},
  \href{http://xxx.lanl.gov/abs/math-ph/0010025}{{\tt math-ph/0010025}}.

\bibitem{Tarasov:1980au}
O.~V. Tarasov, A.~A. Vladimirov, and A.~Yu. Zharkov,
{\it {The Gell-Mann-Low Function of QCD in the Three Loop
                  Approximation}},
 {\em Phys.Lett.} {\bf B93} (1980) 429--432.


\bibitem{Larin:1993tp}
  S.~A.~Larin, J.~A.~M.~Vermaseren,
  {\it {The Three loop QCD Beta function and anomalous dimensions}},
  {\em Phys.\ Lett.} {\bf B303 } (1993) 334--336,
  [\href{http://xxx.lanl.gov/abs/hep-ph/9302208}{{\tt hep-ph/9302208}}].

\bibitem{Vogt:2000ci}
  A.~Vogt,
  {\it {Next-to-next-to-leading logarithmic threshold resummation for deep inelastic scattering and the Drell-Yan process}},
  {\em Phys.\ Lett.} {\bf B497 } (2001) 228--234,
  [\href{http://xxx.lanl.gov/abs/hep-ph/0010146}{{\tt hep-ph/0010146}}].

\bibitem{Catani:2003zt}
  S.~Catani, D.~de Florian, M.~Grazzini, P.~Nason,
  {\it Soft gluon resummation for Higgs boson production at hadron colliders},
  JHEP {\bf 0307 } (2003)  028,
  {\tt [hep-ph/0306211]}.

\end{thebibliography}
\providecommand{\href}[2]{#2}\begingroup\raggedright\endgroup

\end{document}